\documentclass[5p,times]{elsarticle}

\usepackage{lineno,hyperref}
\usepackage{amsmath}
\usepackage{graphicx}
\usepackage{caption}
\usepackage{subcaption}
\graphicspath{ {images/} }
\usepackage{dingbat}

\journal{Optics \& Laser Technology}

\biboptions{sort&compress}
\begin{document}

\begin{frontmatter}

\title{Two-Step Phase Shifting Algorithms: Where Are We?}


\author[CIMAT,UPB]{V\'ictor H. Flores} %\corref{mycorrespondingauthor}}
%\ead{victor.flores@cimat.mx}
\author[CIMAT]{Alan Reyes-Figueroa}
\author[UPB]{C\'esar Carrillo-Delgado}
\author[CIMAT]{Mariano Rivera\corref{mycorrespondingauthor2}}
\ead{mrivera@cimat.mx}
\address[CIMAT]{Centro de Investigaci\'on en Matem\'aticas AC, 36023, Guanajuato, Gto., M\'exico.}
\address[UPB]{Departamento de Ingenier\'ia Rob\'otica, Universidad Polit\'ecnica del Bicentenario, 36283, Silao, Gto., M\'exico.}
\cortext[mycorrespondingauthor2]{Corresponding author}

\begin{abstract}
Two steps phase shifting interferometry has been a hot topic in the recent years. We present a comparison study of 12 representative self--tunning algorithms based on two-steps phase shifting interferometry. We evaluate the performance of such algorithms by estimating the phase step of synthetic and experimental fringe patterns using 3 different normalizing processes: Gabor Filters Bank (GFB), Deep Neural Networks (DNNs) and Hilbert Huang Transform (HHT); in order to retrieve the background, the amplitude modulation and noise. We present the variants of state-of-the-art phase step estimation algorithms by using the GFB and DNNs as normalization preprocesses, as well as the use of a robust estimator such as the median to estimate the phase step. We present experimental results comparing the combinations of the normalization processes and the two steps phase shifting algorithms. Our study demonstrates that the quality of the retrieved phase from of two-step interferograms is more dependent of the normalizing process than the phase step estimation method. 

\end{abstract}


\end{frontmatter}

%\linenumbers

%-------------------------------------------------------------------------------------------------
\section{Introduction}
\label{sec:intro}
%-------------------------------------------------------------------------------------------------

\noindent One of the main challenges in two-steps phase shifting interferometry is the calculation of the phase step, $\delta$. Ideally, the phase of two given interferograms can be calculated by the formula proposed by Muravsky \emph{et al.} \cite{muravsky2011two} if the step between the interferograms is known and the interferograms are normalized with no bakground and amplitud variations. Experimentally, the phase shift is done through optical components such as mirror displacement using a piezoelectric \cite{malacara2007optical}, polarizers \cite{munoz2015measurement}, diffraction grids \cite{rodriguez2008one} or pixelated cameras \cite{kimbrough2006pixelated}. The use of this components in the experimental setup can lead to an erroneous step tuning since if the implementation of the arrangement does not include characterized components. This detuning of the phase displacement causes the presence of harmonics in the recovered phase \cite{servin2014fringe}, for this reason, the Two-Steps Phase Shifting Algorithms (TS--PSAs) consider that the phase step is unknown and include, implicitly or explicitly, its calculation \cite{kreis1992fourier, van1999phase, vargas2012two, vargas2011of, vargas2011st, deng2012two, dalmau2016phase, rivera2016two, kulkarni2018two, flores2019computation, wielgus2015two, luo2015two, meng2008wavefront, farrell1992phase, flores2019robust, liu2015phase, tian2016two, meng2009wavefront}.

The process to obtain the phase in the TS--PSAs consists of 3 stages: 
\begin{enumerate}[a)]
\item Fringe normalization and filter, to overcome the error induced by noise and illumination changes. 
\item Phase step calculation, to avoid the errors caused by detuning.
\item Phase calculation.
\end{enumerate}

In this paper, we present the comparison of different self-tuning algorithms for two-steps phase analysis. We will evaluate the accuracy of the estimated phase step ($\delta$) of each algorithm using couples of synthetic interferograms with different noise levels, random contrast changes and step sizes. Saide \emph{et al.} presented an evaluation of six algorithms and their performance using the Hilbert--Huang Transform as normalization tool of the fringe patterns \cite{saide2017evaluation}. In this study, we will compare eleven up-to-date algorithms against our proposals. The considered algorithms are the following methods: the Fourier based method proposed by Kreis \cite{kreis1992fourier}, the Phase--Step Calibration (PSC) of Van Brug \cite{van1999phase}, the Gram-Schmidt orthonormalization (GS) proposed by Vargas \emph{et al.} \cite{vargas2012two} by using the phase step estimation proposed in Ref \cite{flores2019computation}, the Extreme Value of Interference (EVI) proposed by Deng \emph{et al.} \cite{deng2012two}, the estimation based on Random Points (RP)  proposed by Dalmau  \emph{et al.}  \cite{dalmau2016phase}, the Iterative Robust Estimator (IRE) algorithm proposed by Rivera  \emph{et al.}  \cite{rivera2016two}, the Quadratic Phase Parameter estimation algorithm (QPP) proposed by Kulkarni and Rastogi \cite{kulkarni2018two}, the Tilt--Shift Error estimation proposed by Wielgus \emph{et al.}\cite{wielgus2015two}, the Diamond Diagonal Vectors demodulation algortihm (DDV) proposed by Lou \emph{et al.} \cite{luo2015two}, the General Phase--Shifting Interferometry method of Meng \emph{et al.} \cite{meng2008wavefront} and the Simplified Lissajous Ellipse Fitting by Robust Estimator (SLEF--RE) of Flores and Rivera \cite{flores2019robust}.

Some of the reviewed methods such as GS \cite{vargas2012two} or EVI \cite{deng2012two} use the fringe normalization proposed in \cite{quiroga2003isotropic}. Nevertheless, as proposed in \cite{saide2017evaluation}, an update for these methods is the use of the Hilbert--Huang Transform (HHT) as normalization tool \cite{trusiak2015two, zhang2019two, trusiak2012adaptive, trusiak2014advanced} . On the other hand, according with Refs. \cite{huang2010comparison, zhang2012comparison}, the Windowed Fourier Transform (WTF) based methods have shown to be a robust normalization method, such is the case of the Gabor Filters Bank (GFB), where the kernel of such window is a Gaussian function \cite{daugman1985uncertainty, daugman1988complete, daugman1993high, rivera2016two}. Finally, the DNNs have grown in applications such as de--noising and normalization of interferograms as presented in Refs. \cite{reyes2019deep, zhang2018ffdnet, hao2019batch, yan2019fringe}. Therefore we will evaluate all the algorithms by using these 
three normalization method for preprocessing the FPs. 

The objective of the presented comparison is to point-out which of the stages is more crucial for the phase estimation of the TS--PSAs in order to concentrate the effort of the design of such algorithms. In this work we introduce variations of the presented algorithms in order to improve the performance of the original algorithms. Such variations are based on the implementation of a different pre--filtering process used in the original proposal (such as HHT, GFB or DNNs). Also, we present a modified approach to the solution presented Rivera \emph{et al.} which increases the accuracy of the method by the use of another robust estimator (the median) in the IRE method. Such proposals are included in the comparison here presented.

We organize our paper as follows. We explain the two--step methodology in Section \ref{sec:TS--PSA}. Then, in Section \ref{sec:review}, we present a brief explanation of the evaluated methods identifying the $\delta$ estimation formula or procedure. In Section \ref{sec:proposed_TS--PSA} we introduce the proposed variations that improve the performance of the evaluated methods. In Section \ref{sec:exps}, we will present the analysis of the estimation of the unknown phase step of the TS--PSAs with the HHT, GFB and DNN normalization process at different levels of Gaussian noise with constant step. Then, we calculate the response of each algorithm estimating different phase shifts but with only one normalization process applied to a fixed noise level. For this purpose, we use ten different synthetic patterns. In this section we also present experimental results for prove the feasibility of the proposed algorithms. In Section \ref{sec:dis} we discuss briefly the important aspects of the obtained results and the considerations to be made when using some TS--PSAs. Finally, in Section \ref{sec:con} we present our conclusions.

%-------------------------------------------------------------------------------------------------
\section{ Two-step phase shifting methodology}
\label{sec:TS--PSA}
%-------------------------------------------------------------------------------------------------

\noindent As mentioned before, TS--PSAs consist of three stages: Normalization, step estimation and phase calculation. We start considering the intensity model of  $n$ phase--shifted interferograms:

\begin{equation} \label{eq:image}
I_k(x,y) = a_k(x,y) + b_k(x,y)\cos[\phi(x,y) + d_k] + \eta_k(x,y),
\end{equation}
where $k \in 1,2,\ldots, n$ is the interferogram index, $a_k$ is the background intensity which is different for each interferogram (See \cite{rodriguez2008one,munoz2015measurement} for experimental examples.), $b_k$ is the variable contrast also different for each interferogram, $\phi$ is the phase to be recovered, $d_k$ is the phase step, $\eta_k$ is the noise. For the case of  two--step algorithms we can assume $d_1 = 0$ and $d_2 = \delta$.

In order to normalize the variations of intensity of the background and the contrast, as well as filtering--out the noise, we apply a pre--filtering process such as the ones proposed in \cite{quiroga2003isotropic, trusiak2015two, trusiak2012adaptive, trusiak2014advanced, rivera2016two, reyes2019deep, zhang2018ffdnet, hao2019batch, yan2019fringe} and we obtain the normalized interferograms: 

\begin{align}\label{eq:imagenorm}
	\hat{I}_1(x,y) & = \cos[\phi(x,y)] \\ 
	\hat{I}_2(x,y) & = \cos[\phi(x,y) + \delta]. 
\end{align}

With the filtered images, the phase step can be calculated by using one of the methods presented in section \ref{sec:review}, which will be the objective of our study. 

Finally, using the normalized interferograms and the calculated step, the phase calculation is done by using the equation proposed by Muravsky \emph{et al.} \cite{muravsky2011two}:

\begin{equation}\label{eq:phi}
\phi(x,y) = \arctan\Bigg[\frac{\hat{I}_1(x,y)\cos(\delta)-\hat{I}_2(x,y)} {\hat{I}_1(x,y)\sin(\delta)}\Bigg].
\end{equation}
t
In next section, we will review several methods for estimating the phase step.

%-------------------------------------------------------------------------------------------------
\section{Phase step calculation algorithms}
\label{sec:review}
%-------------------------------------------------------------------------------------------------

\noindent The step calculation is one of the main challenges for the TS--PSAs. In this section we present a brief review of the equations used for the step estimation proposed in each algorithm. Some of the algorithms estimate the phase step as a $\Delta(x,y)$ function. In those cases, we will consider the approximation as the mean value of the phase step map.

%-----------------------------------------------
\subsection{Kreis' Algorithm}
\label{ssec:kreis}
%-----------------------------------------------

\noindent The Kreis algorithm is based on the Fourier transformation. It proposes the use of the sign of the phase step map ($\Delta$) in order to correct the sign ambiguity of closed fringes. Kreis proposes to apply the Fourier transform to each intensity pattern with a width quadrature filter; \emph{i.e.}, Kreis eliminates the low frequencies and one half of the spectra. The phase step is calculated with:

\begin{equation}\label{eq:kreis}
\Delta_{K}(x,y)=\arctan\Bigg[\frac{\Re c_1(x,y) \Im c_2(x,y) - \Im c_1(x,y) \Re c_2(x,y)} {\Re c_1(x,y) \Re c_2(x,y) + \Im c_1(x,y) \Im c_2(x,y)}\Bigg].
\end{equation}
where $c_1$ and $c_2$ are the filtered Fourier transforms of ${I}_1$ and ${I}_2$ respectively, and $\Re$, $\Im$ stand for the real and imaginary parts of the spectrums.

For more details of this method, please refer to \cite{kreis1992fourier}. From this phase difference map $\Delta_K$, we can estimate the  scalar phase step with 

\begin{equation}
\label{eq:deltaK}
\delta_K = \mathbb{E} \{ \Delta_{K} \}
\end{equation}
where $\mathbb{E}\{x\}$ represent the expected (mean) value of $x$.

It is important to remark that the original work does not consider any pre-filtering process at all, which makes it noise--sensitive. As we will demonstrate, we can improve the accuracy of this method by any normalization process. 

%-----------------------------------------------
\subsection{Phase Step Calibration Algorithm (PSC)}
\label{ssec:psc}
%-----------------------------------------------

\noindent The Van Brug's algorithm achieves the Phase Step Calibration (PSC) by using only two images. It is based on computation of the correlation coefficient $\rho$, which is the cosine of an angle between two multidimensional vectors (${I}_1$ and ${I}_2$). As a result, $\rho ({I}_1,{I}_2) = \cos(\delta)$ and $\delta$ can be estimated as:

\begin{equation}\label{eq:vanbrug}
\delta_{PSC}=\arccos\Bigg[\frac{\langle({I}_1-\langle{I}_1\rangle)({I}_2-\langle{I}_2\rangle)\rangle}{\sigma_{\hat{I}_1}\sigma_{\hat{I}_2}}\Bigg],
\end{equation}

\noindent where $\langle \cdot \rangle$ represent the expected (mean) value and $\sigma_{\hat{I}_1}$ and $\sigma_{\hat{I}_2}$ are the standard deviations of the interferograms. 
This method directly computes the  scalar phase step $\delta_{PSC}$. For more details of this method, please refer to \cite{van1999phase}. 

For comparison purposes, we also evaluate the performance of this method by using normalized interferograms, even though the original proposal does not consider any pre-filtering process. 

%-----------------------------------------------
\subsection{Gram-Schmidt orthonormalization Algorithm (GS)}
\label{ssec:gs}
%-----------------------------------------------

\noindent The Gram-Schmidt (GS) orthonormalization method, proposed by Vargas \emph{et al.} calculates the phase between two interferograms with unknown step. The method assumes two previously filtered interferograms, $\hat{I}_1$ and $\hat{I}_2$ (see \eqref{eq:imagenorm}). First, $\hat{I}_1$ is normalized resulting into $\bar{I}_1$. Then, $\hat{I}_2$ is orthogonalized with respect to $\bar{I}_1$ obtaining its projection as $\tilde{I}_2$. Then, $\tilde{I}_2$ is normalized. As result, the orthonormalized pattern $\bar{I}_2$ can be expressed as:

\begin{equation}\label{eq:vargas}
\bar{I}_2(x,y) = -b\sin(\phi(x,y))\sin(\delta),
\end{equation}
where $b$ is the spatially--constant amplitude. Since the patterns are orthogonalized, the phase step cannot be estimated directly, nevertheless, in \cite{flores2019computation} we proposed an algorithm to estimate the phase step by using the GS methodology. 

\begin{equation}
	\label{eq:delta}
 	\delta  _{GS}=  \arcsin \left[ \mathbb{E} \left\{ \frac{\bar I_1(x,y) \tilde{I}_2(x,y)}{ \hat I_1(x,y) \bar I_2(x,y)}  \right\}  \right] 
\end{equation}
where $\mathbb{E}\{x\}$ is the expected value of $x$.
%Thus, $\delta$ can be calculated as the mean value of the of the local phase step map:

%\begin{equation}\label{eq:deltaGS}
%\delta_{GS}= \langle \arctan(\bar{I}_2, \bar{I}_1)  \rangle.
%\end{equation}
For more details, please refer to \cite{vargas2012two, flores2019computation}.

%-----------------------------------------------
\subsection{Extreme Value Interference Algorithm (EVI)}
\label{ssec:evi}
%-----------------------------------------------

\noindent Deng's algorithm uses the Extreme Value Interference (EVI) to compute the phase shift between interferograms. Given the pre-filtered images $\hat{I}_1$ and $\hat{I}_2$, the EVI phase shift map is estimated with: 

\begin{equation}\label{eq:deng1}
\Delta_{EVI}(p)=\arccos\Bigg(\frac{\hat{I}_{2}(p)}{\hat{I}_{1}(p)}\Bigg), \;\;\;\;\; \forall p \in P \cup Q;
\end{equation}
where $P$ is the set of pixel coordinates with maximum local value, $Q$ is set of pixel coordinates with minimum local value, and $p$ are the pixel coordinates that belong to these sets.

The global phase step with:

\begin{equation}\label{eq:deng2}
\delta_{EVI}=\frac{\sum_{p \in P \cup V} \Delta_{EVI}(p)}{\sharp P+ \sharp Q},
\end{equation}
where $\sharp P$ and $\sharp Q$ are the  of the sets $P$ and $Q$, respectively.  For more details, please refer to \cite{deng2012two}.

%-----------------------------------------------
\subsection{Random Points estimation Algorithm (RP)}
\label{ssec:rp}
%-----------------------------------------------

\noindent Dalmau \emph{et al.} proposed a phase estimation algorithm based on computing Random Points (RP) of the interferograms. This consists on extracting pairs of randomly selected pixels from images $\hat{I}_1$ and $\hat{I}_2$. 

Then let $P=[p_1, p_2, \ldots, p_n]$ and $Q = [q_1, q_2, \ldots, q_n]$ be vectors of randomly selected  pixel--coordinates that fulfill $$\hat{I}_1(p_i)\hat{I}_2(p_i) \neq \hat{I}_1(q_i) \hat{I}_2(q_i)$$ for $i=1,2,\ldots, n$. Then, the phase step can be estimated with

\begin{equation}\label{eq:dalmau}
\delta_{RP}=\frac{\sum_{i=1}^{n} a(p_i,q_i) b(p_i,q_i)}{\sum_{p_i,q_i} a(p_i,q_i)^2}
\end{equation}
where $a(p,q)=2[\hat{I}_1(p)\hat{I}_2(p)-\hat{I}_1(q)\hat{I}_2(q)]$ and $b(p,q)=\hat{I}_1(p)^2-\hat{I}_1(q)^2+\hat{I}_2(p)^2-\hat{I}_2(q)^2$. The accuracy of the estimation is improved by increasing the number of samples sampled pixels, $n$. For more details, please refer to \cite{dalmau2016phase}.

%-----------------------------------------------
\subsection{Iterative Robust Estimator Algorithm (IRE)}
\label{ssec:ire}
%-----------------------------------------------

\noindent The GFB algorithm proposed by Rivera \emph{et al.} consists on obtaining the phase of two interferograms and solving the problem of sign ambiguity through an Iterative Robust Estimator (IRE). Considering the individual phases of images $\hat{I}_1$ and $\hat{I}_2$ expressed as $\psi_1$ and $\psi_2$ respectively, from which we can calculate the phase step map as:

\begin{equation}\label{eq:rivera1}
	\Delta_{IRE}=|W[\psi_2-\psi_1]|.
	\end{equation}
where the wrapping operator is defined as $W(z) = z+ 2 \pi n$ with an integer $n$ such that $W(z) \in [-\pi, \pi)$. Then, the global phase shift can, robustly, be computed by iterating

\begin{equation}\label{eq:rivera2}
\delta_{IRE}=\frac{\sum_{x,y}\omega(x,y)\Delta_{IRE}(x,y)}{\sum_{x,y}\omega(x,y)},
\end{equation}
and 
\begin{equation}\label{eq:rivera3}
\omega(x,y)=\frac{\kappa^2}{(\kappa+[\Delta(x,y)-\delta_{IRE}]^2)^2};
\end{equation}
where $\kappa$ is a positive parameter that controls the outlier rejection sensitivity and $\omega$ is the weight; the author proposed  $\kappa=\pi/10$. The initial set is $\omega = 1$ giving the mean of the phase map as the result of the first iteration. For more details, please refer to \cite{rivera2016two}.

%-----------------------------------------------
\subsection{Quadratic Phase Parameter Estimation Algorithm (QPP)}
\label{ssec:qpp}
%-----------------------------------------------

\noindent The QPP algorithm was proposed recently by Kulkarni and Rastogi. They propose the calculation of the phase step by using a quadratic phase approximation from a small window of a pre-filtered interferograms, such as $\hat{I}_1$ and $\hat I_2$. The proposed representation is

\begin{align} \label{eq:kulkarni1}
	\phi_1(\bar x, \bar y) & = c_0 + c_1\bar{x} + c_2\bar{y} + c_3\bar{x}^2 + c_4\bar{xy} + c_5\bar{y}^2 \\
	\phi_2(\bar x, \bar y) & = c_6 + c_1\bar{x} + c_2\bar{y} + c_3\bar{x}^2 + c_4\bar{xy} + c_5\bar{y}^2, 
\end{align}
where $\phi_1$ and $\phi_2$ are the phases of the subsets of $\hat I_1$ and $\hat I_2$ respectively, $\bar{x},\bar{y}$ are the pixel coordinates inside the subset and $c_i$ are the coefficients of the approximation. Since $\phi_2 = \phi_1 + \delta$, then

\begin{equation} \label{eq:kulkarni2}
	\delta = \phi_2 - \phi_1 = c_6 - c_0.
\end{equation}

Kulkarni and Rastogi  proposed the calculation of the coefficients via state space analysis by using the extended Kalman Filter. For more details, please refer to \cite{kulkarni2018two}.

%-----------------------------------------------
\subsection{Tilt--Shift Error estimation Algorithm (TSE)}
\label{ssec:tse}
%-----------------------------------------------

\noindent The TSE algorithm proposed by Wielgus \emph{et al.} consists on the calculation of the phase distribution from two randomly phase shifted interferograms under presence of the tilt-shift error. Assuming two pre--filtered interferograms $\hat I_1$ and $\hat I_2$ where, at least, the background term has been removed, the phase shift map $\Delta_{TSE}$ is given by:

\begin{equation} \label{eq:tse}
	\Delta_{TSE} = \arccos \Bigg\{ \frac{2[\hat{I}_1(x,y)\hat I_2(x,y)] \ast G(x,y)}{[\hat I_1^2(x,y) + \hat I_2^2(x,y)]\ast G(x,y)} \Bigg\}
\end{equation}
where $\ast$ denotes the convolution operator and $G(x,y)$ is the Gaussian mask applied through such convolution. From this phase distribution map $\Delta_{TSE}$, we can estimate the  scalar phase step with 

\begin{equation}
	\label{eq:deltatse}
	\delta_{TSE} = \mathbb{E} \{ \Delta_{TSE} \}
\end{equation}
and $\mathbb{E}\{x\}$ is the expected value of $x$. For more details, please refer to \cite{wielgus2015two}.

%-----------------------------------------------
\subsection{Diamond Diagonal Vectors Algorithm (DDV)}
\label{ssec:ddv}
%-----------------------------------------------

\noindent The DDV algorithm proposes that if two filtered and equally sized interferograms, such as $\hat I_1$ and $\hat I_2$ , satisfy the condition that their norms are almost equal, $\|\hat{I}_1\| \approx \|\hat{I}_2\|$, then, they constitute the adjacent sides of a diamond. Such figure has perpendicular diagonals that can be represented as the sum and the difference of the vectors. The phase shift between the two interferograms can be calculated as:

\begin{equation} \label{eq:ddv}
	\delta_{DDV} = \arctan \Bigg [ \frac{\|\hat{I}_1(x,y) - \hat I_2(x,y)\|} {\|\hat{I}_1(x,y) + \hat I_2(x,y)\|} \Bigg ].
\end{equation}

For more details, please refer to \cite{luo2015two}.

%-----------------------------------------------
\subsection{Generalized Phase Shifting Interferometry Algorithm (GPSI)}
\label{ssec:gpsi}
%-----------------------------------------------

\noindent Meng \emph{et al.} proposed that the GPSI algorithm is able to retrieve the complex phase map of two interferograms without additional preprocessing. Nevertheless, the algorithm considers noiseless interferograms as well as temporally constant background illumination and amplitude. If this criteria is met, such as the case of the use of normalized fringe patterns, the phase shift can be estimated by the use of the generalized quadratic equation

\begin{equation} \label{eq:gpsi}
	\delta_{GPSI} = \arccos \Bigg ( \frac{-B \pm \sqrt{B^2 - 4AC}} {2A} \Bigg )
\end{equation}
where $A = 1$, $B = \langle -2\hat{I}_1\hat{I}_2 \rangle$ and $C = \langle \hat{I}_1^2 + \hat{I}_2^2 - 1 \rangle$. The $\langle \cdot \rangle$ denotes the average value of the resulting term. For more details, please refer to \cite{meng2008wavefront}.

%-----------------------------------------------
\subsection{Simplified Lissajous Ellipse Fitting (SLEF)}
\label{ssec:slef}
%-----------------------------------------------
\noindent The use of the Lissajous figure applied to phase shifting interferometry was originally proposed by Farrell and Player \cite{farrell1992phase}. It consists on using the Lissajous Ellipse Figure (LEF) to represent pixel--wise the intensities of two interferograms with constant amplitude and background terms. Given that the shift between two signals represent the eccentricity of an ellipse, it is possible to obtain such displacement through its conical expression:  

\begin{equation} \label{eq:lef5}
	\theta_1x^2 + \theta_2y^2 + \theta_3x + \theta_4y + \theta_5 = 0 
\end{equation}
where $x = I_1 + I_2$, $y = I_1 - I_2$ and $\theta_i$ are the coefficients of the conical equation of the ellipse. Since the solution can be obtained by solving an overdetermined system, it is proposed the use of the Least Squares method in order to calculate the coefficients of the equation. Given such solution, the phase step is calculated by

\begin{equation} \label{eq:lef}
	\delta_{SLEF} = 2\arctan \Bigg ( \sqrt{ \frac{\theta_1} {\theta_2} } \Bigg ).
\end{equation}
If the case that the Interferograms present spatial and temporal dependency of the background illumination and amplitude term, the normalization process is required. In such case, Flores and Rivera \cite{flores2019robust} proposed the SLEF algorithm which consists on using normalized fringe patterns to calculate the coefficients. Given the normalization process, the conical equation is simplified to

\begin{equation} \label{eq:slef}
	\theta_1x^2 + \theta_2y^2  = 0 
\end{equation}
where only two terms are estimated instead of five. For the estimation of the coefficients, they propose the use of a robust estimator since the pre--filtering process could generate residuals. The phase step between the interferograms is still calculated by \eqref{eq:lef}. For more details, please refer to \cite{farrell1992phase, flores2019robust}.

\begin {table}
\begin{center}
\begin{tabular}{l c l c l}
Algorithm Name					& Acronyms		\\
 \hline
Kreis								& KREIS		\\
Phase Step Calibration				& PSC		\\
Gram--Schmidt orthogonalization		&GS			\\
Extreme Value Interference			&EVI			\\
Random Points estimation				&RP			\\
Iterative Robust Estimator				&IRE			\\
Quadratic Phase Parameter Estimation	&QPP		\\
Tilt--Shift Error estimation				&TSE		\\
Diamond Diagonal Vectors			&DDV		\\
Generalized Phase Shifting Interferometry		&GPSI	\\
Simplified Lissajous Ellipse Fitting		&SLEF		\\
Median Robust Estimator				&MRE		\\
\hline
\end{tabular}
\caption{Algorithms' acronyms}
\label{table:TS--PSAs}
\end{center}
\end{table}

%-------------------------------------------------------------------------------------------------
\section{Proposed variations of TS--PSAs} 
\label{sec:proposed_TS--PSA}
%-------------------------------------------------------------------------------------------------

\noindent For this comparison, our contribution consists on demonstrating the impact of the normalization process of the FPs and the advantages of the use of the GFB as well as the DNNs for such purpose. We consider that algorithms such as Kreis', GS, EVI, PSC, DDV, QPP, GPSI and TSE improve the accuracy of the phase step estimation as well as the quality of the phase map (for clarify purposes, Table \ref{table:TS--PSAs} shows the list of algorithms' acronyms). The original proposals of the algorithms as well as their possible combinations with other normalization are summarized in Table \ref{table:proposals}. The combinations with a check mark ($\ast$) are, in the best of our knowledge, original contributions of this work. In section \ref{sec:exps}, we demonstrate the feasibility of such proposals.

The algorithm proposed by Rivera \emph{et al.} described in subsection \ref{ssec:ire} uses an iterative procedure for the calculation of the phase step. In addition,  we present a variation of the used estimator in order to improve the accuracy of the algorithm at different noise levels. In this case, we propose the use of the Median of the phase map $\Delta_{IRE}$ in \eqref{eq:rivera1} in order to estimate the phase step $\delta_{MRE}$, which denotes the Median Robust Estimator (MRE).

\begin {table}
\begin{center}
\begin{tabular}{l c l c l c | c | c |}
TS--PSAs	 &GFB	&HHT	&DNN	&Other/None \\
 \hline
KREIS	&$\ast$	&\cite{saide2017evaluation}	&$\ast$	&\cite{kreis1992fourier}			\\
PSC		&$\ast$	&$\ast$	&$\ast$	&\cite{van1999phase}		\\
GS		&$\ast$	&\cite{trusiak2015two, saide2017evaluation}	&$\ast$	&\cite{vargas2012two}		\\
EVI		&$\ast$	&\cite{saide2017evaluation, zhang2019two}	&$\ast$	&\cite{deng2012two}		\\
RP		&\cite{dalmau2016phase}	&$\ast$	&$\ast$	&\cite{dalmau2016phase}		\\
IRE		&\cite{rivera2016two}	&NA		&NA		\\
QPP		&$\ast$	&$\ast$	&$\ast$	&\cite{kulkarni2018two}		\\
TSE		&$\ast$	&\cite{wielgus2015two, saide2017evaluation}	&$\ast$		\\
DDV		&$\ast$	&$\ast$	&$\ast$	&\cite{luo2015two}		\\
GPSI	&$\ast$	&$\ast$	&$\ast$	&\cite{meng2008wavefront}		\\
SLEF	&\cite{flores2019robust}	&\cite{liu2016simultaneous}	&$\ast$	&\cite{farrell1992phase}		\\
MRE		&$\ast$	&NA		&NA		\\
\hline
\end{tabular}
\caption{References of the evaluated methods. ($\ast$ non-previously reported).}
\label{table:proposals}
\end{center}
\end{table}

%-------------------------------------------------------------------------------------------------
\section{Experiments and results}
\label{sec:exps}
%-------------------------------------------------------------------------------------------------

%----------------------------------------
\subsection{Constant step, variable noise level}
%----------------------------------------

\begin{figure}[ht]
	\centering	
	\begin{subfigure}[ht]{0.19\linewidth}
		\centering
		\includegraphics[width=1\linewidth]{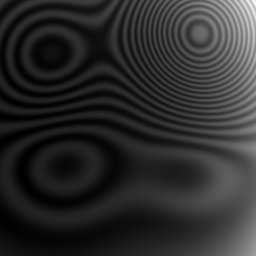}
		\caption{\centering $\sigma=0$}
		\label{fig:n01}
	\end{subfigure}
	\begin{subfigure}[ht]{0.19\linewidth}
		\centering
		\includegraphics[width=1\linewidth]{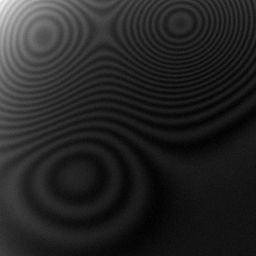}
		\caption{\centering $\sigma=0.25$}
		\label{fig:n11}
	\end{subfigure}
	\begin{subfigure}[ht]{0.19\linewidth}
		\centering
		\includegraphics[width=1\linewidth]{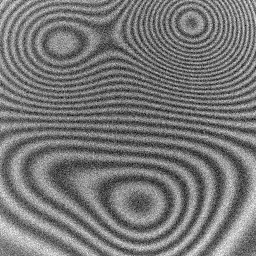}
		\caption{\centering  $\sigma=0.5$}
		\label{fig:n21}
	\end{subfigure}
	\begin{subfigure}[ht]{0.19\linewidth}
		\centering
		\includegraphics[width=1\linewidth]{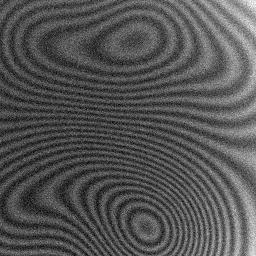}
		\caption{\centering  $\sigma=0.75$}
		\label{fig:n31}
	\end{subfigure}
	\begin{subfigure}[ht]{0.19\linewidth}
		\centering
		\includegraphics[width=1\linewidth]{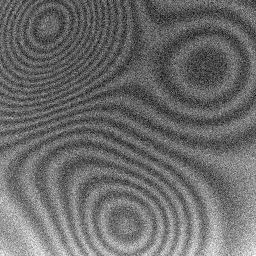}
		\caption{\centering  $\sigma=1.0$}
		\label{fig:n41}
	\end{subfigure}
	
	\begin{subfigure}[ht]{0.19\linewidth}
		\centering
		\includegraphics[width=1\linewidth]{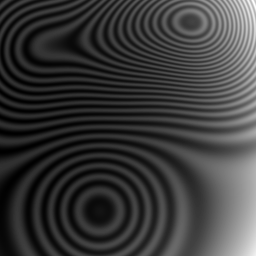}
		\caption{\centering  $\sigma=0$}
		\label{fig:n02}
	\end{subfigure}
	\begin{subfigure}[ht]{0.19\linewidth}
		\centering
		\includegraphics[width=1\linewidth]{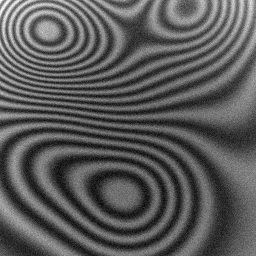}
		\caption{\centering  $\sigma=0.25$}
		\label{fig:n12}
	\end{subfigure}
	\begin{subfigure}[ht]{0.19\linewidth}
		\centering
		\includegraphics[width=1\linewidth]{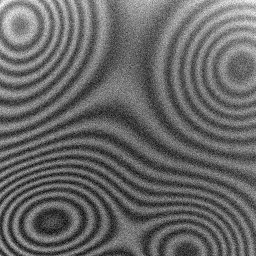}
		\caption{\centering  $\sigma=0.5$}
		\label{fig:n22}
	\end{subfigure}
	\begin{subfigure}[ht]{0.19\linewidth}
		\centering
		\includegraphics[width=1\linewidth]{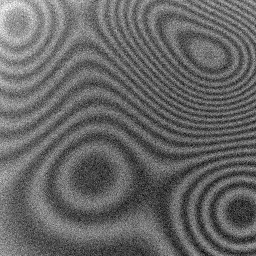}
		\caption{\centering  $\sigma=0.75$}
		\label{fig:n32}
	\end{subfigure}
	\begin{subfigure}[ht]{0.19\linewidth}
		\centering
		\includegraphics[width=1\linewidth]{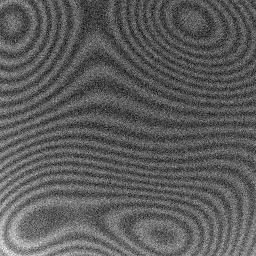}
		\caption{\centering  $\sigma=1.0$}
		\label{fig:n42}
	\end{subfigure}
	
	\caption{Synthetic patterns.}
	\label{fig:patterns}
\end{figure}

\noindent The evaluation of the algorithms was done with ten different sets of synthetic fringe patterns of $1024 \times 1024$ pixels, represented as in \eqref{eq:image} with variable background illumination components and amplitude modulations as well as five different Gaussian noise levels $(\sigma = [0.0, 0.25, 0.5, 0.75, 1.0])$.  The phase step between the two patterns was $\pi/3$. Figure \ref{fig:patterns} shows the different FPs used for the experiment. It is important to remark that all the images presented the different noise levels with variable backgrounds and amplitudes, nevertheless, for illustrative purposes, we present each pattern with different noise levels. In this case Figures \ref{fig:n01} and \ref{fig:n02} the fringe patterns without noise but the background variations and the amplitude modulations are noticeable. Figures \ref{fig:n21} and \ref{fig:n22} show the patterns with a noise level of $\sigma = 0.5$. It is important to notice that the spatial and temporal variations of the background and the amplitude are still present. Figures \ref{fig:n41} and \ref{fig:n42} show a high noise level of $\sigma = 1.0$ where the background variation is not noticeable at all because of the noise level. 

The methods to be evaluated are the ones presented in Section \ref{sec:review} including the MRE proposal of Section \ref{sec:proposed_TS--PSA} giving a total of 12 methods. The different TS--PSAs names will be abbreviated  as shown in Table \ref{table:TS--PSAs}. In order to carry out a fair comparison, we implemented the algorithm with three different pre--filtering processes: GFB, DNN and HHT. We evaluated the accuracy of the phase step estimation of each method using these techniques. In Table \ref{table:times} we present the computational time required by each of the normalization process applied on every FP. All the algorithms were tested on a 3.4 Hz Intel Core i5 computer with 32GB. The GFB was implemented in Python while the HHT pre?filtering was performed via the EFEMD method available for Matlab. Also, it is important to note that the DNN normalization is computed in a GPU NVIDIA GTX 1080ti, which can be applicable to the filtering process after a training process that takes around 47 min.

\begin {table}
\begin{center}
\begin{tabular}{l c l c l}
Normalization Process			& Time (sec)		\\
 \hline
Gabor Filters Bank (GFB)			& 6.64		\\
Hilbert--Huang Transform (HHT)	& 0.39		\\
Deep Neural Network (DNN)		& 0.12		\\
\hline
\end{tabular}
\caption{Computational costs of the normalization processes}
\label{table:times}
\end{center}
\end{table}

%-----------------------------------------------
\subsubsection{Normalization by Gabor Filters Bank (GFB)}
\label{sec:gfbpre}
\begin{figure}[ht]
	\centering	
	\begin{subfigure}[ht]{0.19\linewidth}
		\centering
		\includegraphics[width=1\linewidth]{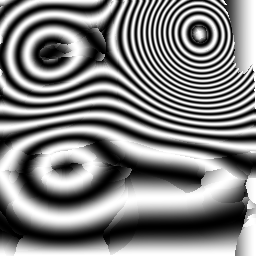}
		\caption{\centering $\sigma=0$}
	\end{subfigure}
	\begin{subfigure}[ht]{0.19\linewidth}
		\centering
		\includegraphics[width=1\linewidth]{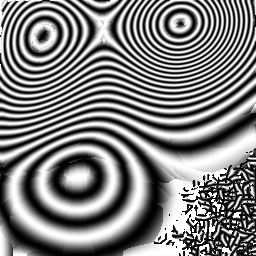}
		\caption{\centering $\sigma=0.25$}
	\end{subfigure}
	\begin{subfigure}[ht]{0.19\linewidth}
		\centering
		\includegraphics[width=1\linewidth]{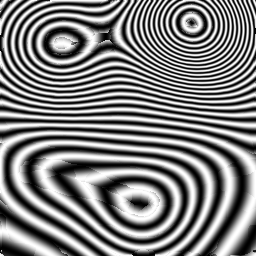}
		\caption{\centering  $\sigma=0.5$}
	\end{subfigure}
	\begin{subfigure}[ht]{0.19\linewidth}
		\centering
		\includegraphics[width=1\linewidth]{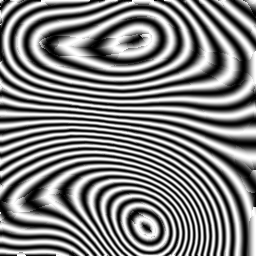}
		\caption{\centering  $\sigma=0.75$}
	\end{subfigure}
	\begin{subfigure}[ht]{0.19\linewidth}
		\centering
		\includegraphics[width=1\linewidth]{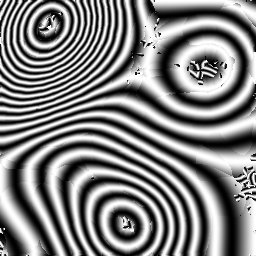}
		\caption{\centering  $\sigma=1.0$}
	\end{subfigure}
	
	\begin{subfigure}[ht]{0.19\linewidth}
		\centering
		\includegraphics[width=1\linewidth]{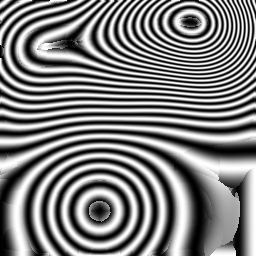}
		\caption{\centering  $\sigma=0$}
	\end{subfigure}
	\begin{subfigure}[ht]{0.19\linewidth}
		\centering
		\includegraphics[width=1\linewidth]{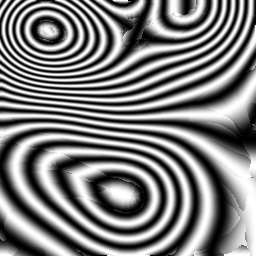}
		\caption{\centering  $\sigma=0.25$}
	\end{subfigure}
	\begin{subfigure}[ht]{0.19\linewidth}
		\centering
		\includegraphics[width=1\linewidth]{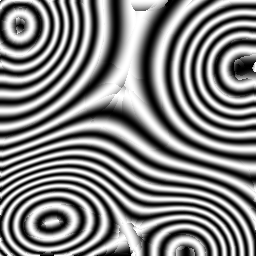}
		\caption{\centering  $\sigma=0.5$}
	\end{subfigure}
	\begin{subfigure}[ht]{0.19\linewidth}
		\centering
		\includegraphics[width=1\linewidth]{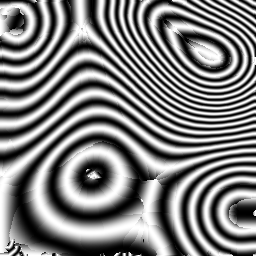}
		\caption{\centering  $\sigma=0.75$}
	\end{subfigure}
	\begin{subfigure}[ht]{0.19\linewidth}
		\centering
		\includegraphics[width=1\linewidth]{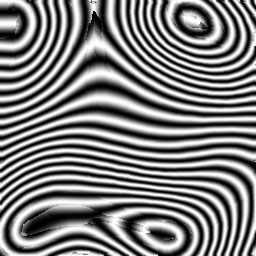}
		\caption{\centering  $\sigma=1.0$}
	\end{subfigure}
	
	\caption{Synthetic patterns normalized with GFB}
	\label{fig:patterns_gfb}
\end{figure}

%-----------------------------------------------
\noindent As first test, we present the comparison of the phase step calculation of the previously mentioned TS--PSAs. In Figure \ref{fig:patterns_gfb} we present the resulting normalized patterns which correspond to each of the figures shown in Figure \ref{fig:patterns}. It is clear that the GFB have trouble with low frequency fringes but it is robust to high noise levels.

\begin{figure}[ht]
    \centering
    \includegraphics[width=1\linewidth]{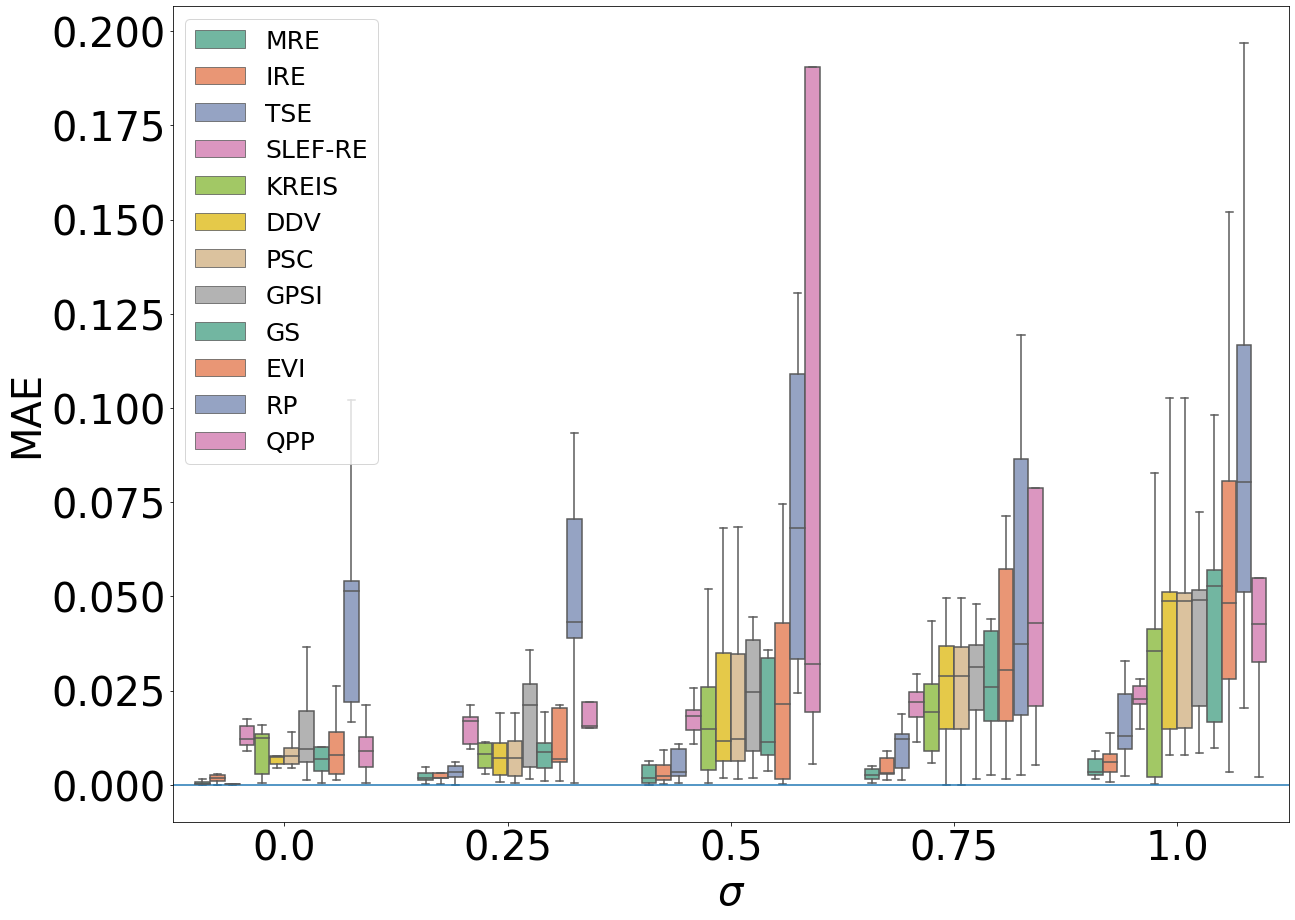}
    \caption{MAE distribution of the algorithms at different levels of noise with GFB normalization.}
    \label{fig:all_gfb}
\end{figure}

\begin{figure}[ht]
    \centering
    \includegraphics[width=1\linewidth]{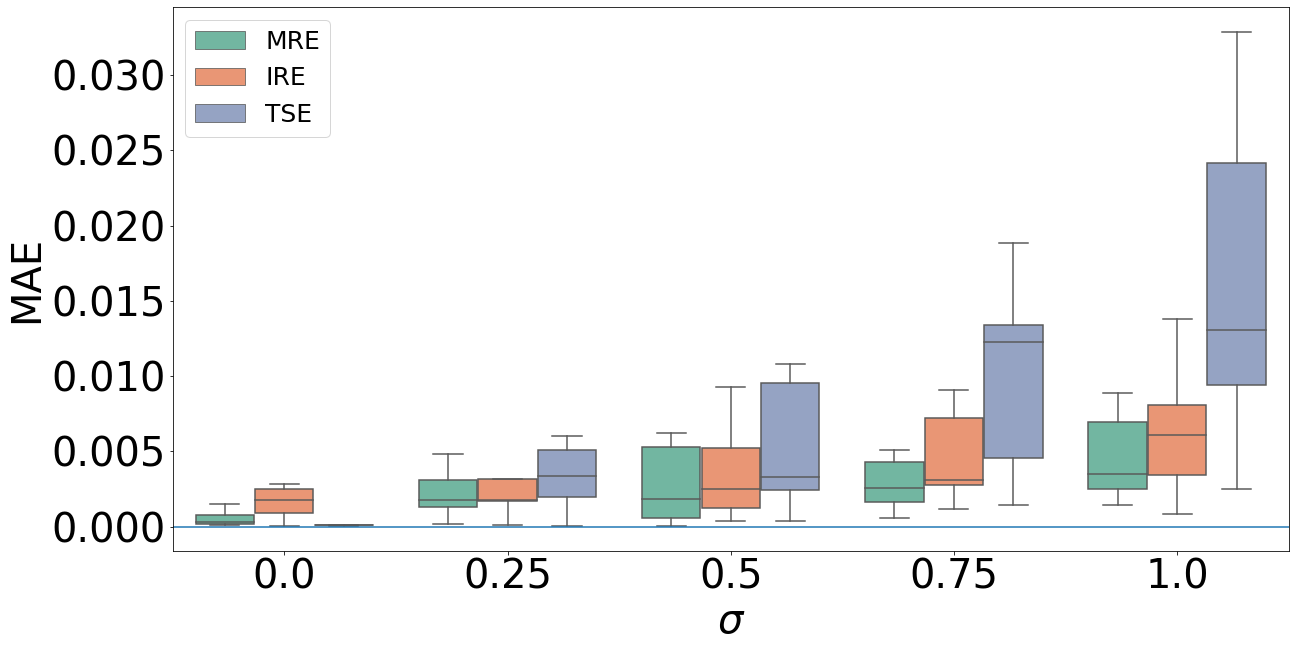}
    \caption{MAE distribution of the MRE, IRE and TSE methods using GFB normalized patterns.}
    \label{fig:rk}
\end{figure}

The results of the evaluation are shown in Figure \ref{fig:all_gfb}. For each noise level, the Mean Absolute Error (MAE) of the step estimation of each algorithm is shown with its respective deviation. Some interesting results are that the robust version of the SLEF algorithm (SLEF--RE) presented good stability but with some deviation. Kreis, DDV, PSC, GPSI, GS and EVI presented very similar behavior in all noise levels. Finally, the RP algorithm as well as the QPP presented high variance and low accuracy among the test. The RP is quiet inestable since the random nature of the estimation of the step, because it could consider some residuals of the filtered patterns. On the other hand, the QPP algorithm is highly inestable since it depends of the chosen window of evaluation, This will be discussed in detail in Section \ref{sec:dis}. Finally, the MRE and the IRE algorithms had the best accuracy and less variance among the algorithms no matter the noise level. Also, the TSE algorithm proved to be best algorithms of the ones that use a simple formula. Figure \ref{fig:rk} depicts in detail, the behavior of the these TS--PSAs.

%As mentioned in section \ref{ssec:RK}, our proposal is to improve the results of the Kreis' method by using the phases obtained through GFB for computing the local phases and the use of a median estimator. As a result, Figure \ref{fig:rk} presents the MAE of the proposed RK method and we compare it with the MRE, IRE and TSE algorithms, which presented the less error in the previous comparison. 

%It is obvious that both, the MRE and the RK algorithms present the same behavior, both have an error less than $0.005 rad$ with low variance. In Section \ref{sec:dis} we present a brief discussion of such results.

%-----------------------------------------------
\subsubsection{Normalization by Deep Neural Networks (DNNs)}
\label{ssec:dnnpre}
%-----------------------------------------------

\begin{figure}[ht]
	\centering	
	\begin{subfigure}[ht]{0.19\linewidth}
		\centering
		\includegraphics[width=1\linewidth]{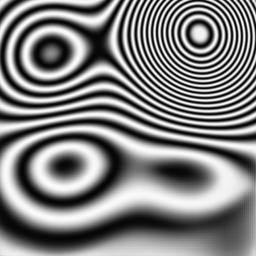}
		\caption{\centering $\sigma=0$}
	\end{subfigure}
	\begin{subfigure}[ht]{0.19\linewidth}
		\centering
		\includegraphics[width=1\linewidth]{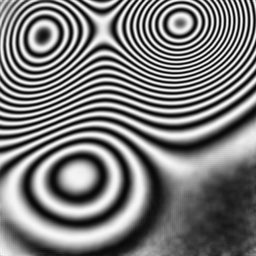}
		\caption{\centering $\sigma=0.25$}
	\end{subfigure}
	\begin{subfigure}[ht]{0.19\linewidth}
		\centering
		\includegraphics[width=1\linewidth]{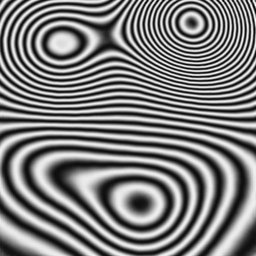}
		\caption{\centering  $\sigma=0.5$}
	\end{subfigure}
	\begin{subfigure}[ht]{0.19\linewidth}
		\centering
		\includegraphics[width=1\linewidth]{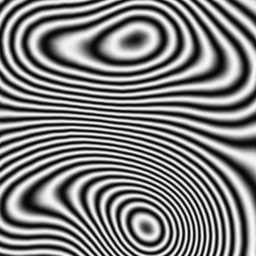}
		\caption{\centering  $\sigma=0.75$}
	\end{subfigure}
	\begin{subfigure}[ht]{0.19\linewidth}
		\centering
		\includegraphics[width=1\linewidth]{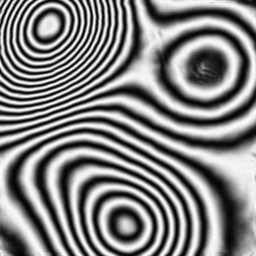}
		\caption{\centering  $\sigma=1.0$}
	\end{subfigure}
	
	\begin{subfigure}[ht]{0.19\linewidth}
		\centering
		\includegraphics[width=1\linewidth]{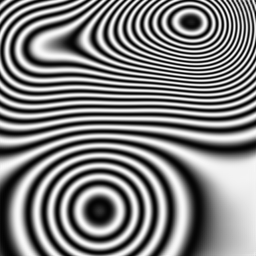}
		\caption{\centering  $\sigma=0$}
	\end{subfigure}
	\begin{subfigure}[ht]{0.19\linewidth}
		\centering
		\includegraphics[width=1\linewidth]{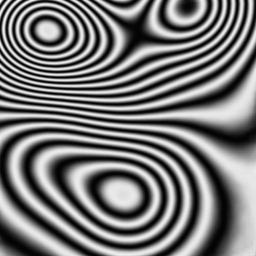}
		\caption{\centering  $\sigma=0.25$}
	\end{subfigure}
	\begin{subfigure}[ht]{0.19\linewidth}
		\centering
		\includegraphics[width=1\linewidth]{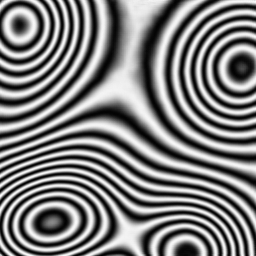}
		\caption{\centering  $\sigma=0.5$}
	\end{subfigure}
	\begin{subfigure}[ht]{0.19\linewidth}
		\centering
		\includegraphics[width=1\linewidth]{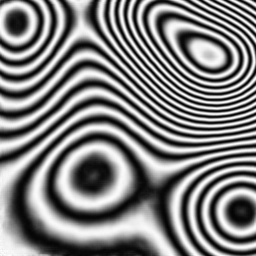}
		\caption{\centering  $\sigma=0.75$}
	\end{subfigure}
	\begin{subfigure}[ht]{0.19\linewidth}
		\centering
		\includegraphics[width=1\linewidth]{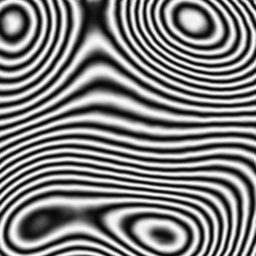}
		\caption{\centering  $\sigma=1.0$}
	\end{subfigure}
	
	\caption{Synthetic patterns normalized with DNNs}
	\label{fig:patterns_dnn}
\end{figure}

\begin{figure}[ht]
    \centering
    \includegraphics[width=1\linewidth]{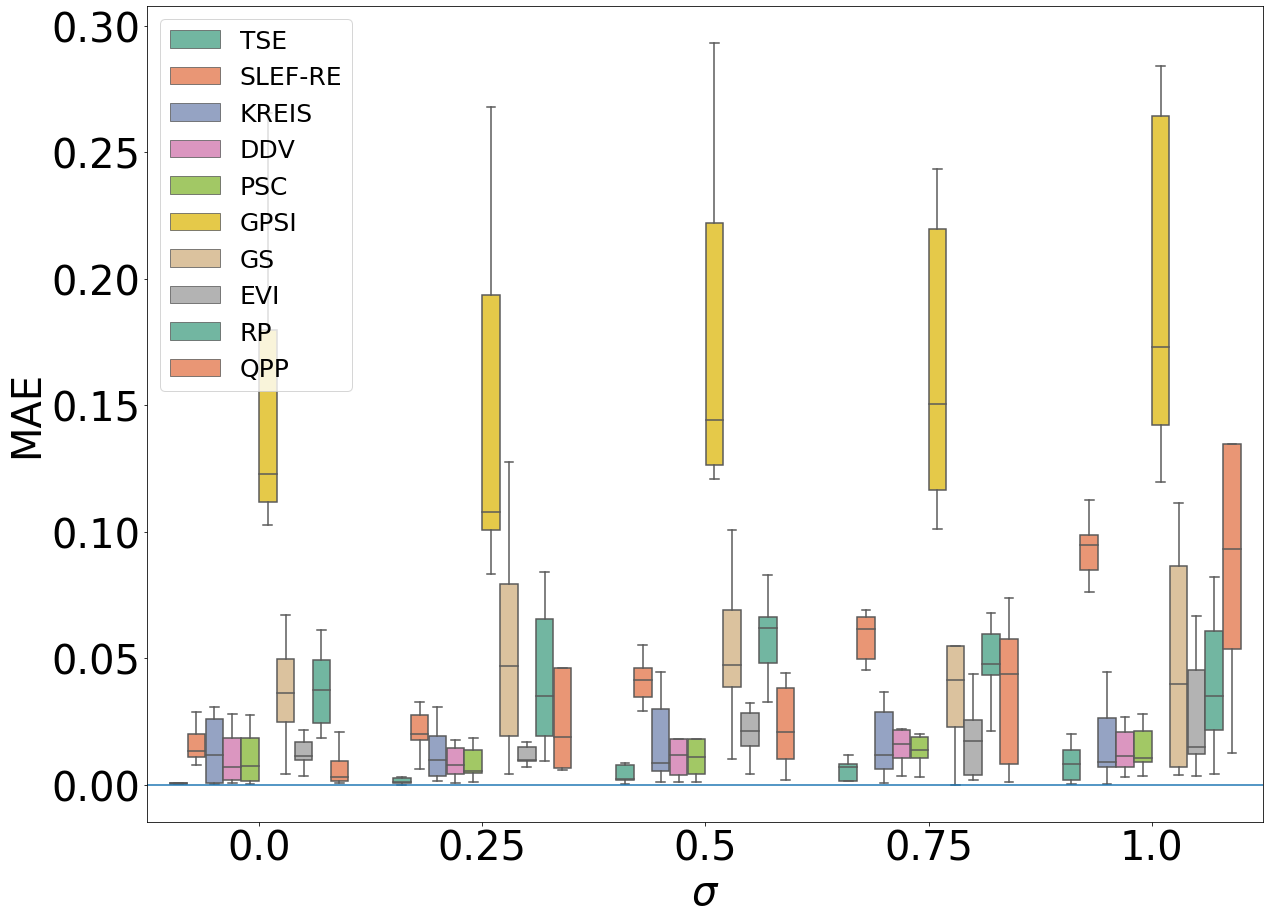}
    \caption{MAE distribution of the algorithms at different levels of noise with DNN normalization.}
    \label{fig:all_dnn}
\end{figure}

\begin{figure}[ht]
    \centering
    \includegraphics[width=1\linewidth]{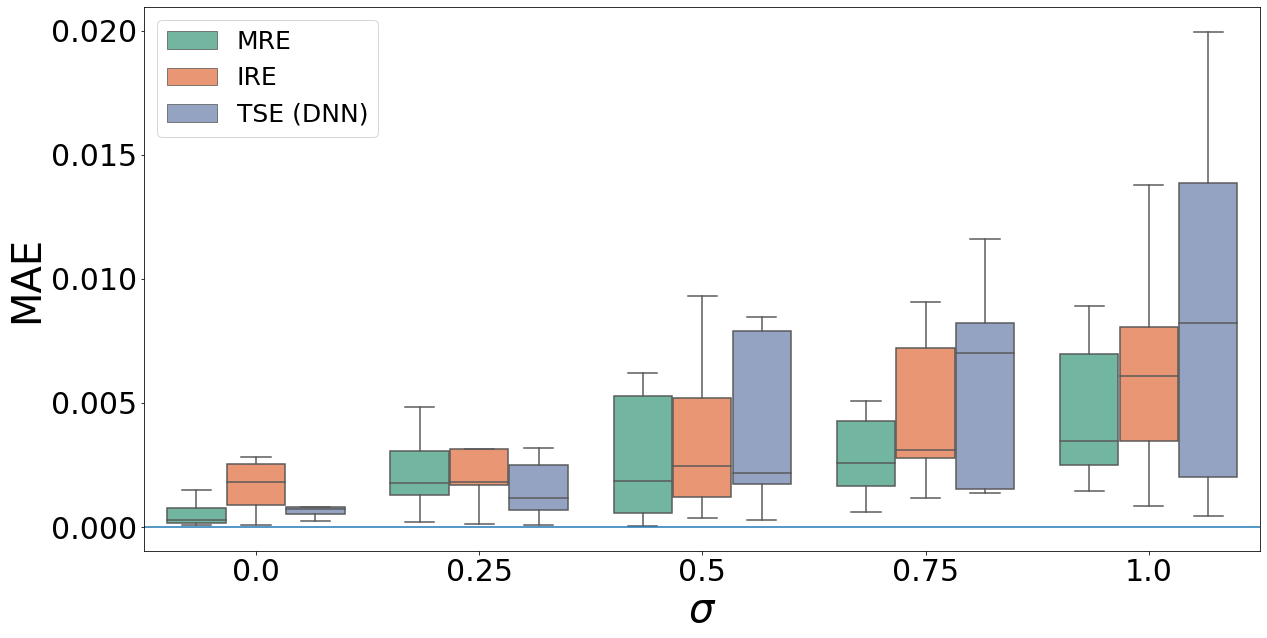}
    \caption{MAE distribution of the MRE, IRE and TSE algorithms. The TSE uses patterns with DNN normalization.}
    \label{fig:rk_dnn}
\end{figure}

\noindent We present the implementation of the pre--filtering process by using the Deep Neural Network (DNN) proposed by Rivera and Reyes \cite{reyes2019deep}. An example of these normalized patterns is shown in Figure \ref{fig:patterns_dnn}. These patterns correspond to the normalized version of the ones presented in Figure \ref{fig:patterns}. The obtained results are presented in Figures \ref{fig:all_dnn} and \ref{fig:rk_dnn}.

In Figure \ref{fig:all_dnn} we can observe that the phase step estimation of the algorithms using DNN normalized FPs. It can be seen that almost the same for the most of the algorithms. It is important to clarify that the IRE and MRE algorithms did not used DNN normalized patterns since they are completely based on the use of GFBs. On the other hand, the QPP method improved drastically its estimation since the DNN normalization presents less discontinuities with respect of the GFB normalization. Moreover, the GPSI presented a bias term as well as high variability compared with its results in the GFB normalization.

In Figure \ref{fig:rk_dnn} we can see in detail the performance of the most accurate algorithm, in this case TSE, with respect to the GFB based ones IRE and MRE. It is important to remark that with a DNN normalization, the TSE algorithm proposed by Wielgus \cite{wielgus2015two} presented the best accuracy and stability at different noise levels.

%-----------------------------------------------
\subsubsection{Normalization by Hilbert--Huang Transform (HHT)}
\label{ssec:hhtpre}
%----------------------------------------------

\begin{figure}[ht]
	\centering	
	\begin{subfigure}[ht]{0.19\linewidth}
		\centering
		\includegraphics[width=1\linewidth]{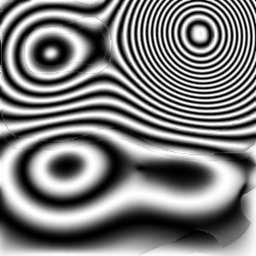}
		\caption{\centering $\sigma=0$}
	\end{subfigure}
	\begin{subfigure}[ht]{0.19\linewidth}
		\centering
		\includegraphics[width=1\linewidth]{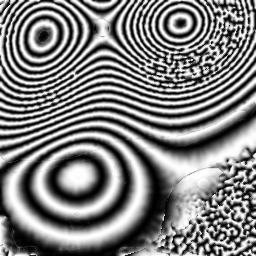}
		\caption{\centering $\sigma=0.25$}
	\end{subfigure}
	\begin{subfigure}[ht]{0.19\linewidth}
		\centering
		\includegraphics[width=1\linewidth]{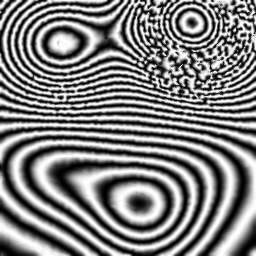}
		\caption{\centering  $\sigma=0.5$}
	\end{subfigure}
	\begin{subfigure}[ht]{0.19\linewidth}
		\centering
		\includegraphics[width=1\linewidth]{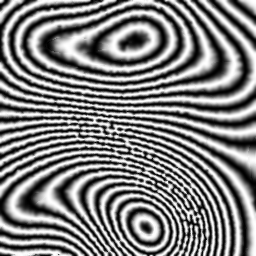}
		\caption{\centering  $\sigma=0.75$}
	\end{subfigure}
	\begin{subfigure}[ht]{0.19\linewidth}
		\centering
		\includegraphics[width=1\linewidth]{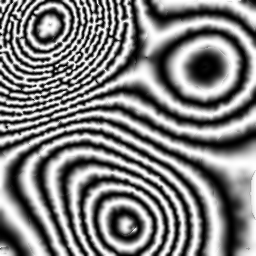}
		\caption{\centering  $\sigma=1.0$}
	\end{subfigure}
	
	\begin{subfigure}[ht]{0.19\linewidth}
		\centering
		\includegraphics[width=1\linewidth]{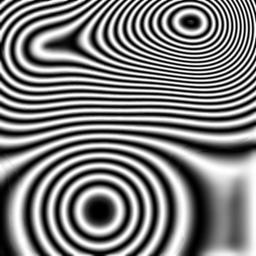}
		\caption{\centering  $\sigma=0$}
	\end{subfigure}
	\begin{subfigure}[ht]{0.19\linewidth}
		\centering
		\includegraphics[width=1\linewidth]{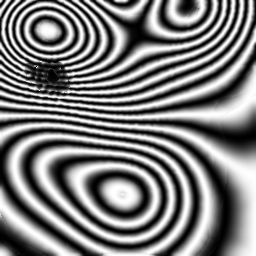}
		\caption{\centering  $\sigma=0.25$}
	\end{subfigure}
	\begin{subfigure}[ht]{0.19\linewidth}
		\centering
		\includegraphics[width=1\linewidth]{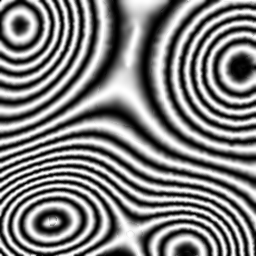}
		\caption{\centering  $\sigma=0.5$}
	\end{subfigure}
	\begin{subfigure}[ht]{0.19\linewidth}
		\centering
		\includegraphics[width=1\linewidth]{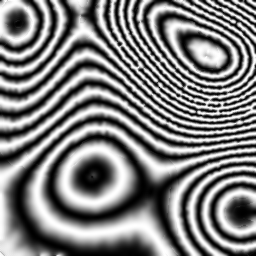}
		\caption{\centering  $\sigma=0.75$}
	\end{subfigure}
	\begin{subfigure}[ht]{0.19\linewidth}
		\centering
		\includegraphics[width=1\linewidth]{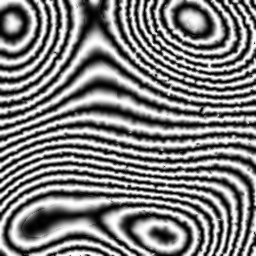}
		\caption{\centering  $\sigma=1.0$}
	\end{subfigure}
	
	\caption{Synthetic patterns normalized with HHT}
	\label{fig:patterns_hht}
\end{figure}

\noindent We present the estimation of the phase step of of the TS--PSAs using the Hilbert--Huang Transform (HHT) normalization, as proposed by Trusiak \emph{et al.} \cite{trusiak2014advanced}. A sample of the resulting normalized patterns is shown in Figure \ref{fig:patterns_hht}, each of them corresponding to the ones presented in Figure \ref{fig:patterns}. 

\begin{figure}[ht]
    \centering
    \includegraphics[width=1\linewidth]{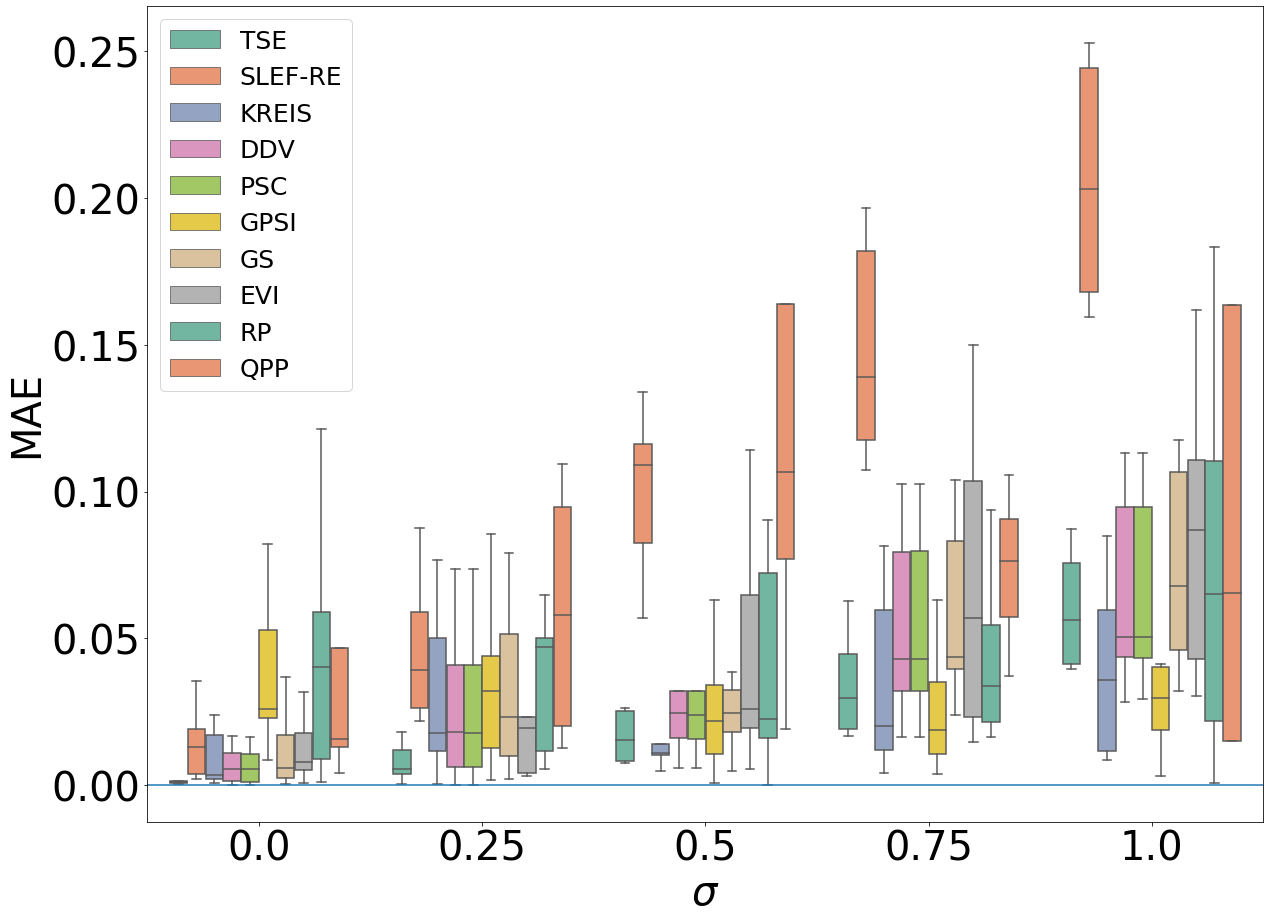}
    \caption{MAE distribution of the algorithms at different levels of noise with HHT normalization.}
    \label{fig:all_hht}
\end{figure}

In Figure \ref{fig:all_hht} we present the results of the evaluation of the phase step estimation of the different TS--PSAs with HHT normalized patterns. In this case, it is noticeable that the best accuracy was presented by the TSE algorithm. On the other hand, all the TS--PSAs presented a slight increase in error due to the higher presence of noise. The most affected TS--PSAs are the SLEF--RE and the QPP as the noise increases. It is important to mention that, for visualization purposes, we did not include some fliers from the QPP method, which were estimations around $7 rad$.

\begin{figure}[ht]
    \centering
    \includegraphics[width=1\linewidth]{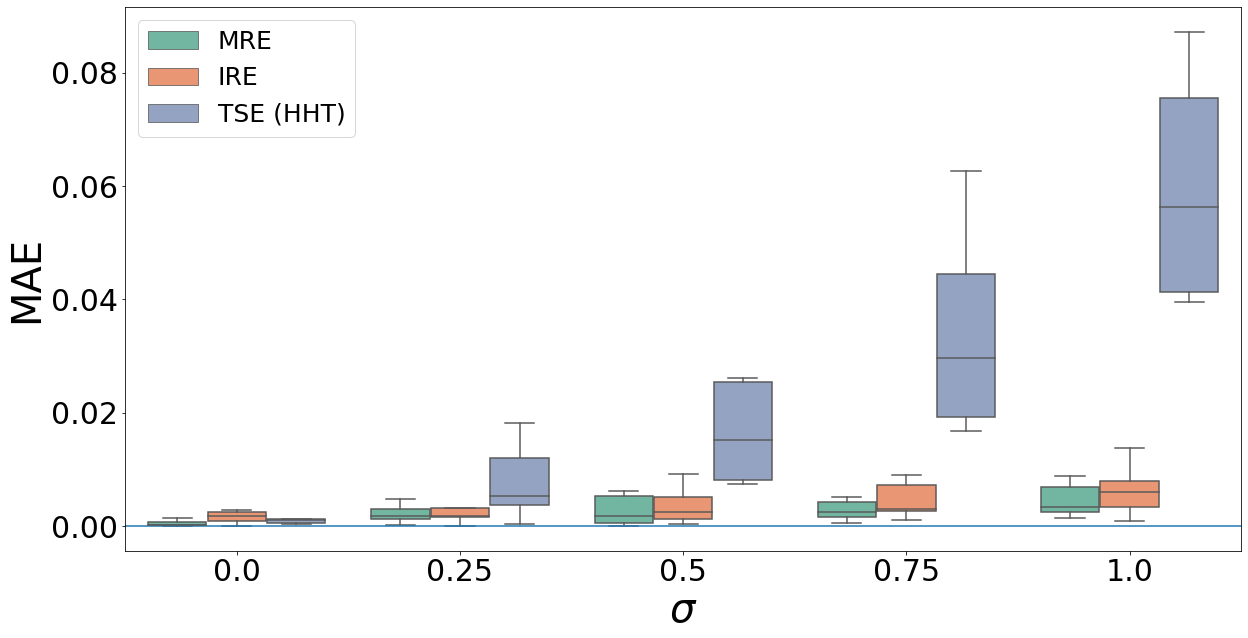}
    \caption{MAE distribution of the MRE, IRE and TSE algorithms. The TSE uses patterns with HHT normalization.}
    \label{fig:rk_hht}
\end{figure}

Figure \ref{fig:rk_hht} depicts in detail the performance of the TSE algorithm with HHT normalization with respect to the GFB based algorithms, IRE and MRE.

%-----------------------------------------------
\subsection{Variable step, constant noise level}
\label{sec:shift}
%-----------------------------------------------

\begin{figure}[ht]
    \centering
    \includegraphics[width=1\linewidth]{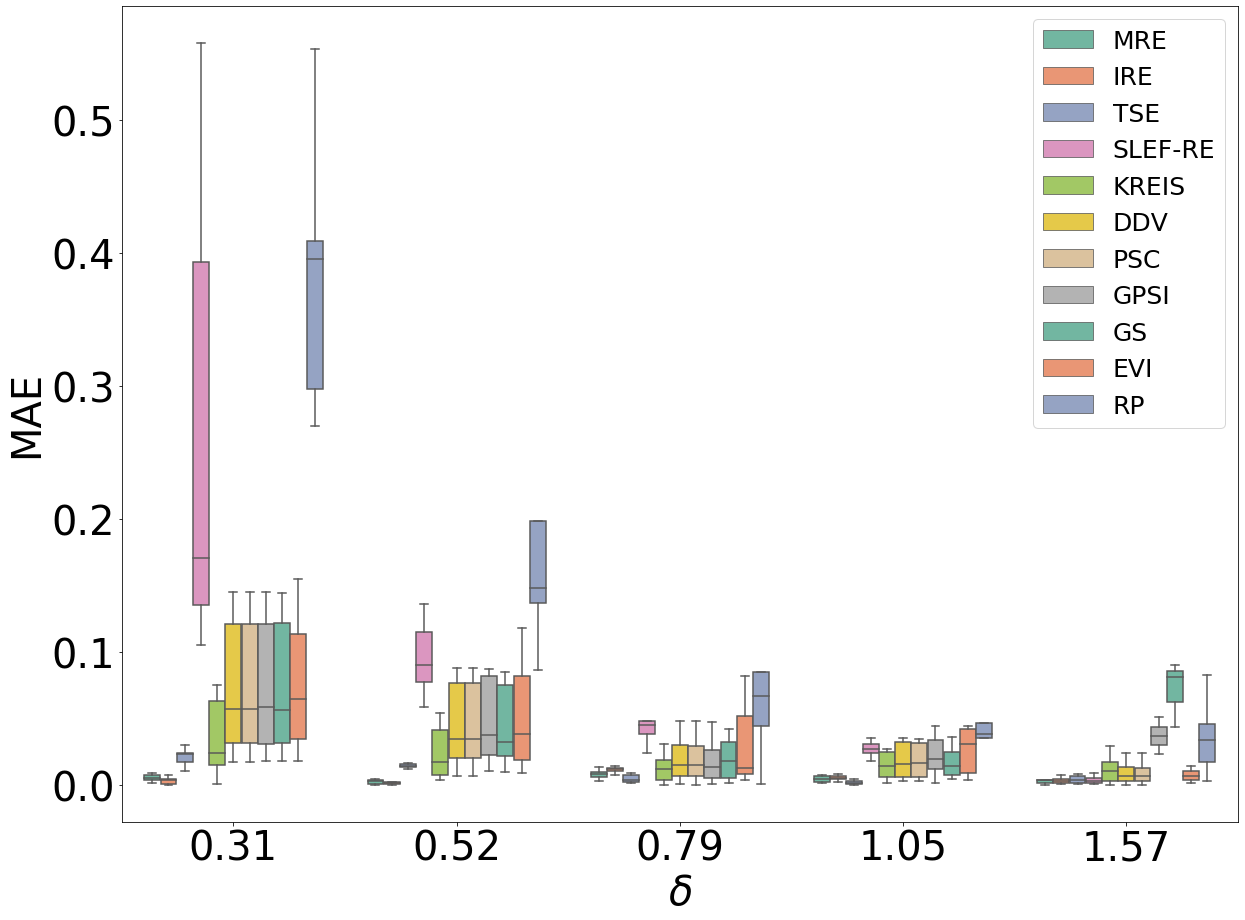}
    \caption{MAE distribution of the step estimation. Constant noise level $\sigma=0.5$}
    \label{fig:error_shift}
\end{figure}

\noindent For this analysis, we considered the same ten patterns with a constant noise level of $\sigma=0.5$ and phase steps of $$\delta = [\pi/10, \pi/6, \pi/4, \pi/3, \pi/2]$$ which are some of the most representative values. The steps of $\delta > \pi/2$ are no considered since they behave the same. The results of the estimation are shown in Figure \ref{fig:error_shift} where each algorithm was tested with a pre--filtered fringe pattern by the GFB method. Again, in this test the MRE, IRE and TSE algorithms presented the best results in all steps. The SLEF--RE and the RP algorithms presented high variability and low accuracy if the step is small, but they improve as the step gets bigger. The non robust version of  Kreis' algorithm also maintained a error lower to $0.05 rad$. The DDV, PSC, GPSI, GS and EVI algorithms behave the same among the different steps. This MAE distribution shows that all of the algorithms are good estimators for steps of $\pi/2$ and values near to it.

\begin{figure}[ht]
    \centering
    \includegraphics[width=1\linewidth]{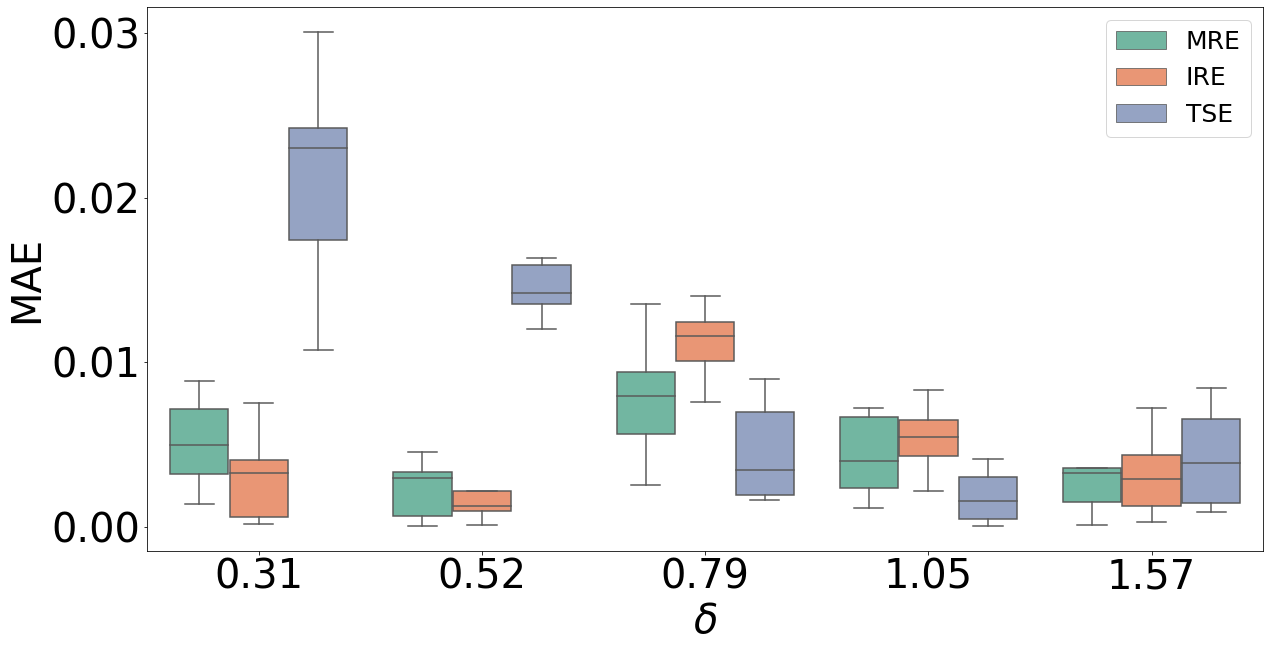}
    \caption{MAE distribution of the step estimation. Constant noise level $\sigma=0.5$}
    \label{fig:rk_shift}
\end{figure}

The RK algorithm also present the tendency of low MAE on the estimation of steps near to $\pi/2$ as shown in Figure \ref{fig:rk_shift}.

%-----------------------------------------------
\subsection{Phase error maps, synthetic patterns}
\begin{figure*}[ht]
    \centering
    \includegraphics[width=1\linewidth]{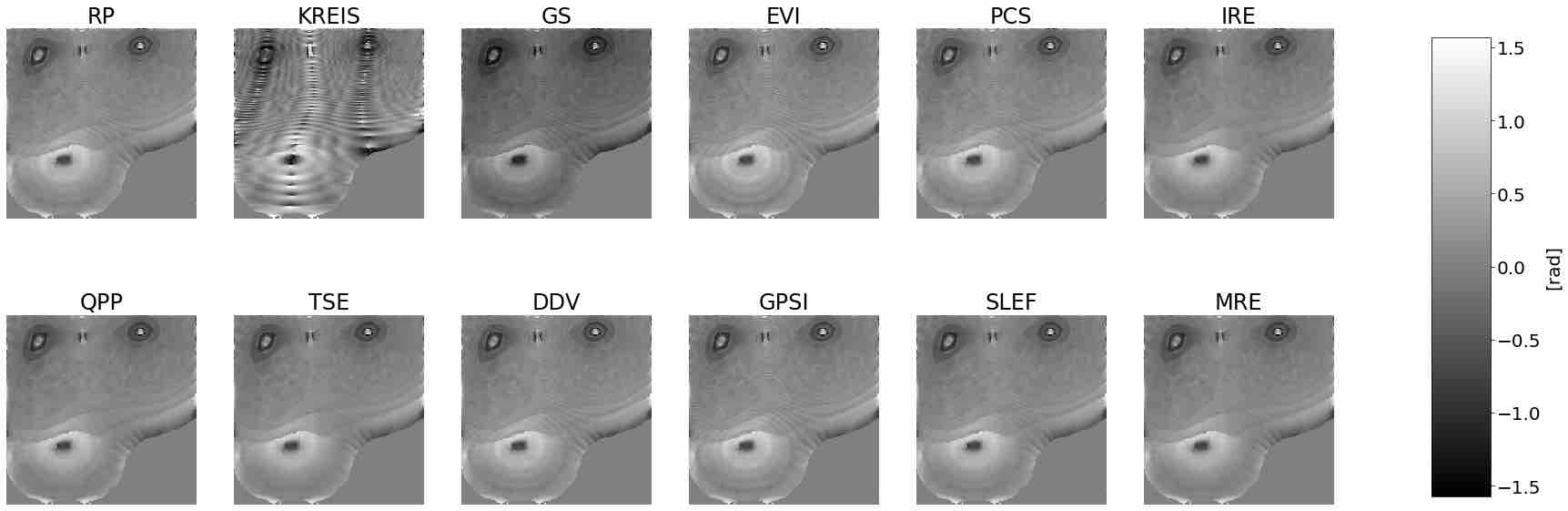}
    \caption{Phase error maps with GFB normalization.}
    \label{fig:error_gfb}
\end{figure*}

\begin{figure*}[ht]
    \centering
    \includegraphics[width=1\linewidth]{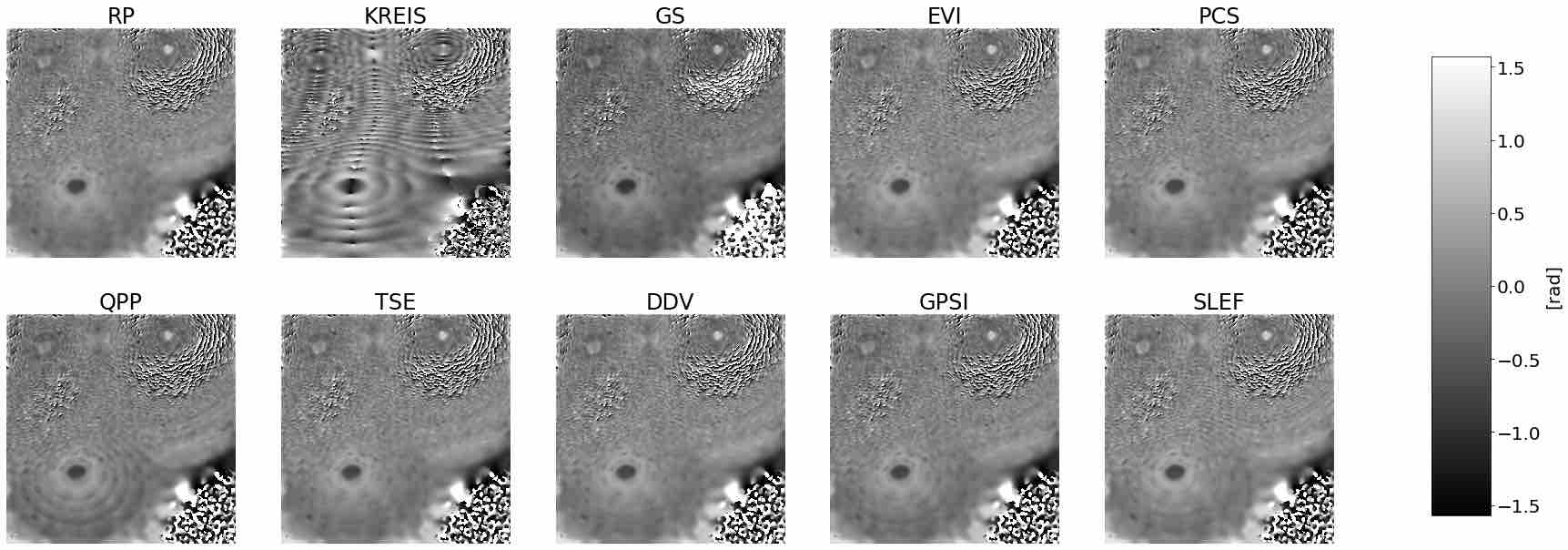}
    \caption{Phase error maps with HHT normalization.}
    \label{fig:error_hht}
\end{figure*}

\begin{figure*}[ht]
    \centering
    \includegraphics[width=1\linewidth]{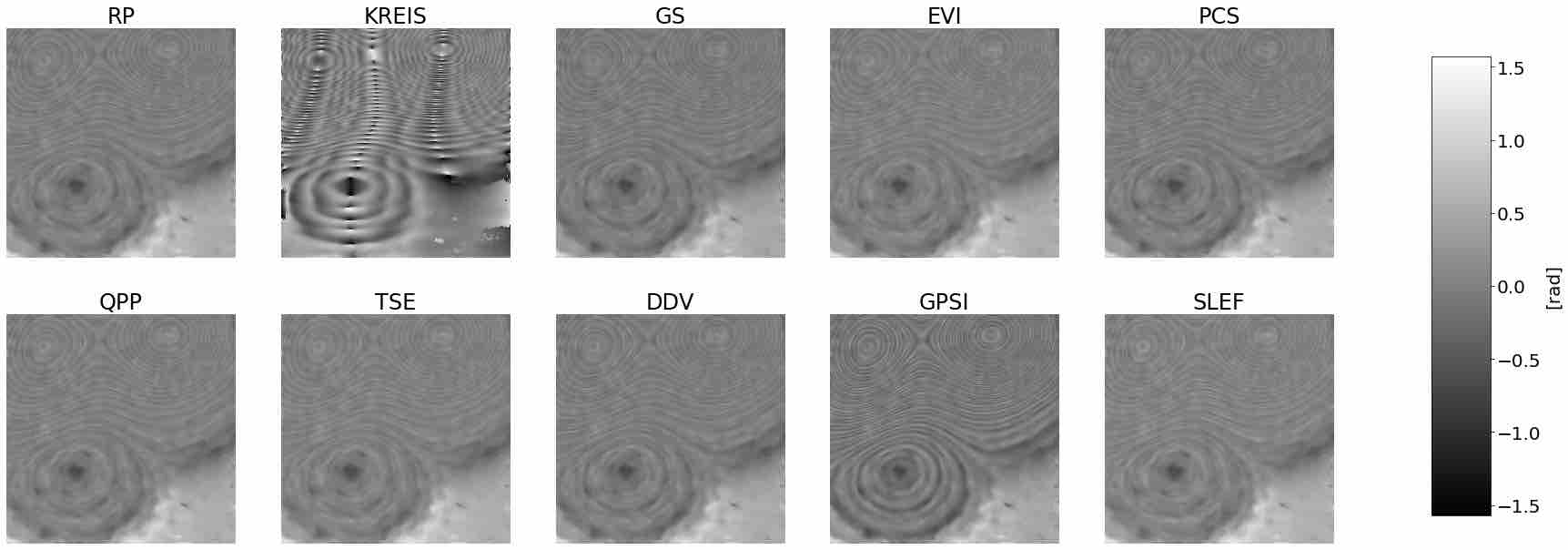}
    \caption{Phase error maps with DNN normalization.}
    \label{fig:error_dnn}
\end{figure*}

%-----------------------------------------------
\noindent In this section we present a comparison of errors in the estimated phase map in order to demonstrate the impact of the normalization process and the used TS--PSAs. Figures \ref{fig:error_gfb}, \ref{fig:error_hht} and \ref{fig:error_dnn} present the phase map errors of the estimated phases using GFB, HHT and DNN normalizations with TS--PSAs described in Section \ref{sec:TS--PSA}. 

The estimated phases correspond to the synthetic FP shown in Figure \ref{fig:n11} with a noise level of $\sigma = 0.5$. First, the FPs were normalized with the different methods previously mentioned. Then, the phase step, as well as the phase map, were estimated with the TS--PSAs here presented. Finally, each phase map was subtracted to the ideal wrapped phase map, and the obtained error map was wrapped in order to show the harmonics generated due to the phase step estimation.

In Figure \ref{fig:error_gfb} we present the 12 evaluated methods using the GFB pre--filtering process. It can be seen that the low frequency regions of the FP present artifacts introduced by the normalization process. The range of the phase error maps are between $-\pi/2$ and $\pi/2$. Kreis' algorithm presents the most notable variation, due to the spectral filter used in the algorithm.

Figure \ref{fig:error_hht} shows the phase error maps of HHT normalized patterns. In this case we present 10 of the methods, since IRE and MRE are GFB based methods. Even thought the error oscillates around zero, it is visible that the noise is not completely filtered out at FP's regions with low and very high frequencies. 

The phase error maps obtained with the DNN normalized FPs are presented in Figure \ref{fig:error_dnn}. In this case, the error maps are smoother than the previously presented ones, this is mainly because the DNNs present higher robustness to noise, nevertheless, it requires the necessary training data to consider all fringe cases, contrary to the GFB and HHT processes.

%-----------------------------------------------
\subsection{Optical experimental data}
\begin{figure}[ht]
	\begin{subfigure}[ht]{0.24\linewidth}
		\centering
		\includegraphics[width=1\linewidth]{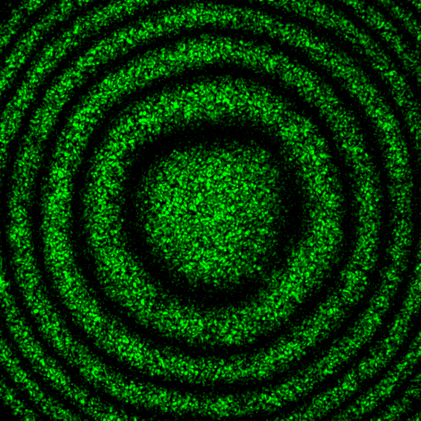}
		\caption{$\delta=0$}
		\label{fig:I1}	
	\end{subfigure}
	\begin{subfigure}[ht]{0.24\linewidth}
		\centering
		\includegraphics[width=1\linewidth]{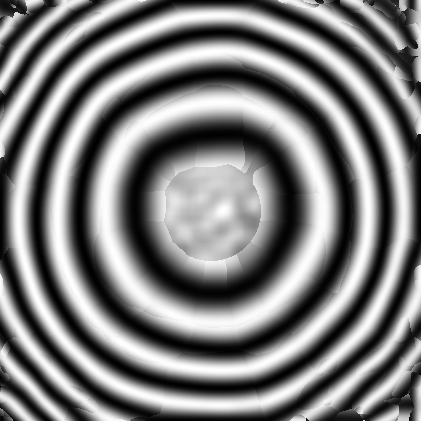}
		\caption{$\hat{I}_1$, GFB}
		\label{fig:GB1}	
	\end{subfigure}
	\begin{subfigure}[ht]{0.24\linewidth}
		\centering
		\includegraphics[width=1\linewidth]{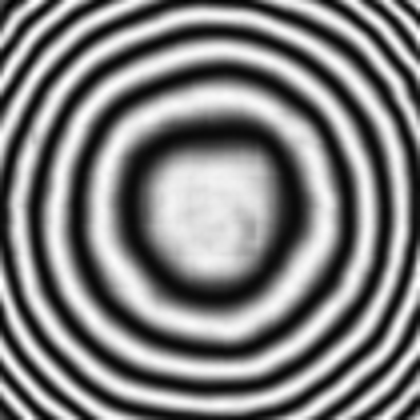}
		\caption{$\hat{I}_1$, DNN}
		\label{fig:DNN1}	
	\end{subfigure}
	\begin{subfigure}[ht]{0.24\linewidth}
		\centering
		\includegraphics[width=1\linewidth]{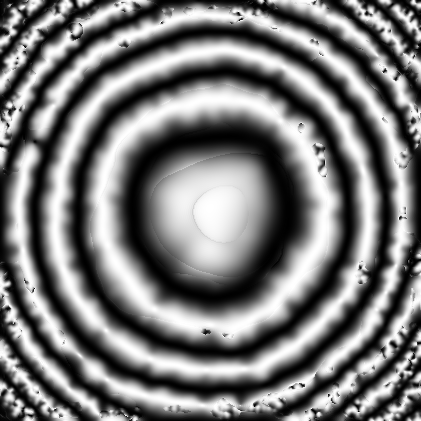}
		\caption{$\hat{I}_1$, HHT}
		\label{fig:HHT1}	
	\end{subfigure}
	
	\begin{subfigure}[ht]{0.24\linewidth}
		\centering
		\includegraphics[width=1\linewidth]{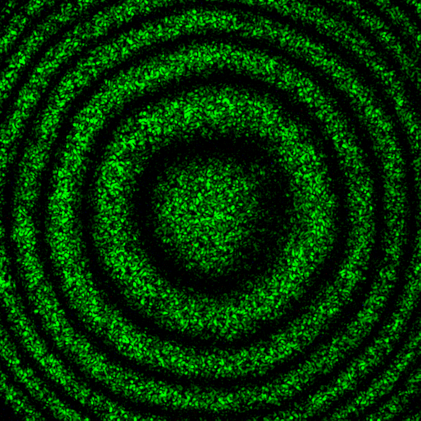}
		\caption{$\delta=\pi/3$}
		\label{fig:I2}	
	\end{subfigure}
	\begin{subfigure}[ht]{0.24\linewidth}
		\centering
		\includegraphics[width=1\linewidth]{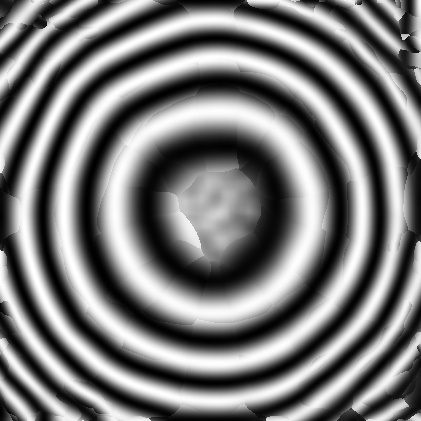}
		\caption{$\hat{I}_2$, GFB}
		\label{fig:GB2}	
	\end{subfigure}
	\begin{subfigure}[ht]{0.24\linewidth}
		\centering
		\includegraphics[width=1\linewidth]{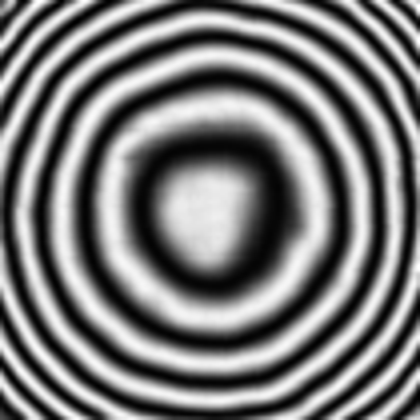}
		\caption{$\hat{I}_2$, DNN}
		\label{fig:DNN2}	
	\end{subfigure}
	\begin{subfigure}[ht]{0.24\linewidth}
		\centering
		\includegraphics[width=1\linewidth]{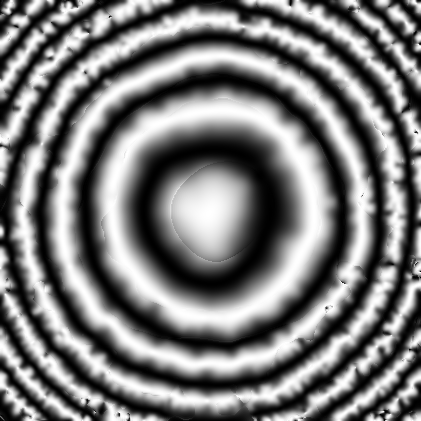}
		\caption{$\hat{I}_2$, HHT}
		\label{fig:HHT2}	
	\end{subfigure}
	\caption{Experimental results of a Polarizing Cyclic Path Interferometer (PCPI).}
	\label{fig:experiment}
\end{figure}

%-----------------------------------------------
\noindent We present the analysis of the response of each algorithm with experimental FPs with different normalization processes applied. We obtained such FPs by the implementation of a Polarizing Cyclic Path Interferometer (PCPI) as the one proposed in \cite{toto20174d} in its radial mode. We used a $\lambda = 532 nm$ laser to illuminate the sample and the reference arm of the interferometer. The image was captured with a $2048\times1536$ CCD camera. To generate the phase shift between the patterns, we rotated the polarizer an angle of $\pi/6$ which induces a displacement of the fringes of $\delta = \pi/3 \approx1.047 rad$, such displacement was performed with a calibrated rotational mount and a polarizer (for more details of the arrangement refer to \cite{toto20174d}). Figures \ref{fig:I1} and \ref{fig:I2} present the obtained experimental FPs.

For the evaluation of the TS--PSAs presented in section \ref{sec:review}, we implemented 3 different normalization methods which are shown in Figure \ref{fig:experiment}.

Figures \ref{fig:GB1} and \ref{fig:GB2} present the normalized patterns using the Gabor Filters Bank (GFB) proposed in \cite{rivera2016two, rivera2018robust}. We implemented the filters considering 8 orientations ($\theta_k = k\pi/8$ for $k=0,1,2,\ldots,7$) and four frequencies corresponding to the periods of pixel size $\tau = \{20, 35, 45, 55\}$.

In Figures \ref{fig:DNN1} and \ref{fig:DNN2} we show the normalized patterns using Deep Neural Networks as proposed in \cite{reyes2019deep, zhang2018ffdnet, hao2019batch, yan2019fringe}. In this case, we present the implementation of the V--Net proposed in \cite{reyes2019deep}.

Finally, Figures \ref{fig:HHT1} and \ref{fig:HHT2} present the results of the normalization using the Hilbert--Huang Transform (HHT) proposed in \cite{saide2017evaluation, trusiak2012adaptive, trusiak2014advanced, bhuiyan2008fast}. For our results, we implemented the Enhaced Fast Empirical Mode Decomposition (EFEMD) \cite{trusiak2014advanced}.

Several of the evaluated algorithms assume that the normalization of the FPs is performed by the algorithm proposed in \cite{quiroga2003isotropic}, nevertheless, as mentioned in \cite{saide2017evaluation}, the HHT presents better results for fringe normalization, such is that,  several state-of-the-art proposals have been done such as Refs. \cite{trusiak2015two, liu2016simultaneous, zhang2019two} (Refer to Table \ref{table:proposals}). For this reason, we start our comparison with the HHT normalization and compare its results with the GFB and DNN methods.

%-----------------------------------------------
\subsubsection{Variable step, experimental patterns}
%-----------------------------------------------

\noindent Now, we use the optical experimental data corresponding to fringe patterns presented in Figure \ref{fig:experiment}. The representative steps were $$\delta = [\pi/4, \pi/3, \pi/2, 2\pi/3, 3\pi/4].$$ 

\begin{figure}[ht]
		\centering
		\includegraphics[width=1\linewidth]{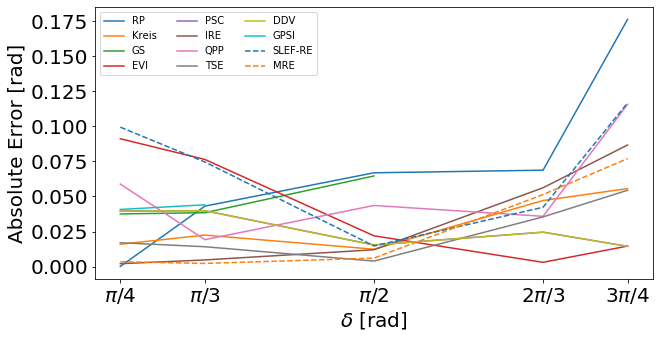}
		\caption{Absolute error of the step estimation with experimental interferograms. GFB normalization.}
		\label{fig:exp_shift_gfb}
\end{figure}

\begin{figure}[ht]
		\centering
		\includegraphics[width=1\linewidth]{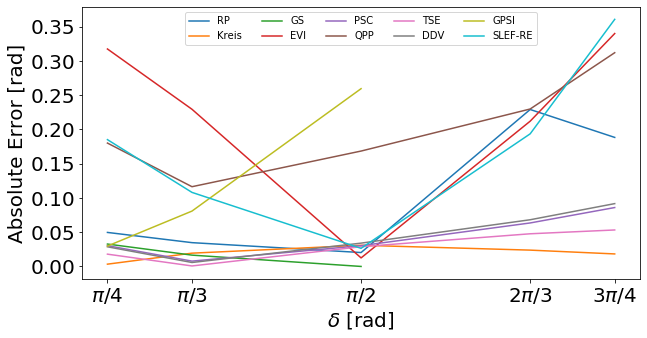}
		\caption{Absolute error of the step estimation with experimental interferograms. HHT normalization.}
		\label{fig:exp_shift_hht}
\end{figure}

\begin{figure}[ht]
		\centering
		\includegraphics[width=1\linewidth]{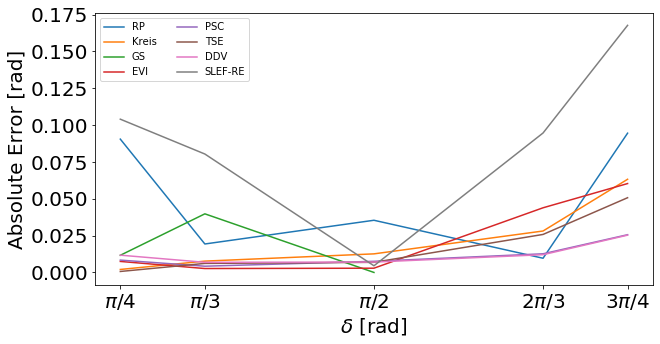}
		\caption{Absolute error of the step estimation with experimental interferograms. DNN normalization.}
		\label{fig:exp_shift_dnn}
\end{figure}

Figures \ref{fig:exp_shift_gfb}, \ref{fig:exp_shift_hht} and \ref{fig:exp_shift_dnn} present the graphs of absolute error of the phase step estimations using the TS--PSAs presented in Section \ref{sec:TS--PSA} (we suggest the reader to refer to the digital version in order to visualize the graph's colors). 

Figure \ref{fig:exp_shift_gfb} shows the absolute error obtained with the GFB normalized patterns. It can be seen that the error is minimum at a $\pi/2$ phase shift, while it grows as it tends to the extreme values (closer to zero or $\pi$). The GS and GPSI algorithms present higher error than the range of the graphic in steps larger than $\pi/2$, thus, they are omitted in the graph. 

The error associated to HHT normalized patterns is presented in Figure \ref{fig:exp_shift_hht}, here we can see that most of the algorithms present high error in the estimation of the phase step in the extreme values. This could be associated to the noisy results of the normalizing process. The IRE and MRE algorithms are not included given that they are GFB based.

Finally, in Figure \ref{fig:exp_shift_dnn} we present errors in the estimation of the TSPAs using DNN normalized patterns. In general, all the algorithms present errors minor than 0.1. In this case, the QPP presented errors higher than the range of the graph, which is associated to the evaluation region used in the algorithm (more details are discussed in Section \ref{sec:dis}. The GPSI algorithm presented errors higher too, which do not allow to visualize the accuracy of the other algorithms correctly.

%-----------------------------------------------
\subsubsection{Phase map, experimental patterns}
%-----------------------------------------------

\begin{figure*}[ht]
	\centering
	\begin{tabular}{lccc}
		\hline
		Algorithm&
		GFB&
		DNN&
		HHT\\
		\hline
		MRE&
		\begin{subfigure}[ht]{0.15\linewidth}
			\includegraphics[width=1\linewidth]{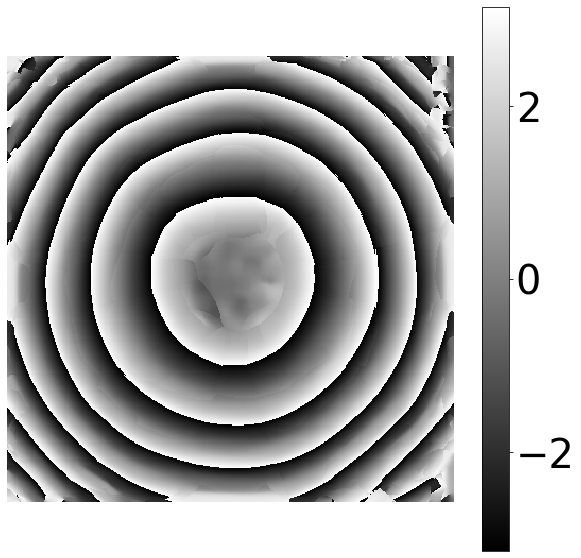}
			\caption{$\delta = 1.044rad$}
			\label{fig:mre_gfb}	
		\end{subfigure}&NA&NA\\

		IRE&
		\begin{subfigure}[ht]{0.15\linewidth}
			\includegraphics[width=1\linewidth]{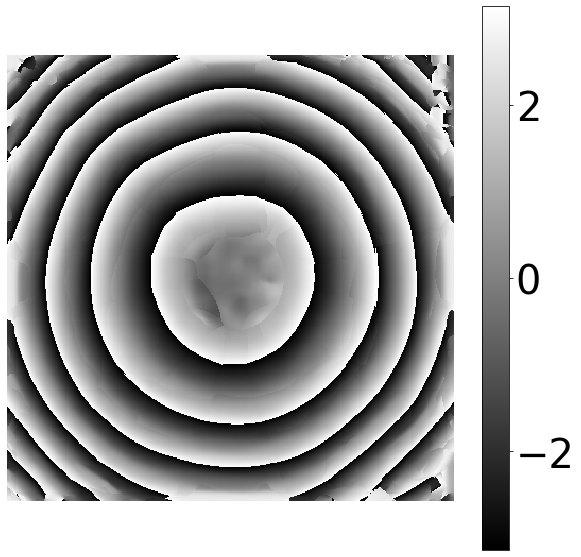}
			\caption{$\delta = 1.051 rad$}
			\label{fig:ire_gfb}	
		\end{subfigure}&NA&NA\\

		TSE&
		\begin{subfigure}[ht]{0.15\linewidth}
			\includegraphics[width=1\linewidth]{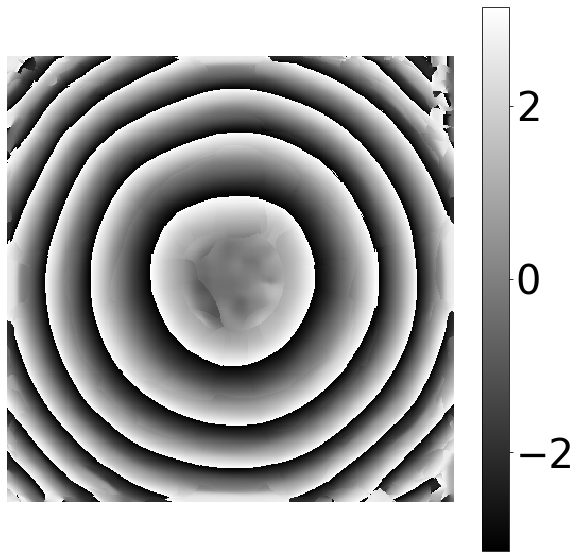}
			\caption{$\delta = 1.089 rad$}
			\label{fig:tse_gfb}	
		\end{subfigure}&
		\begin{subfigure}[ht]{0.15\linewidth}
			\includegraphics[width=1\linewidth]{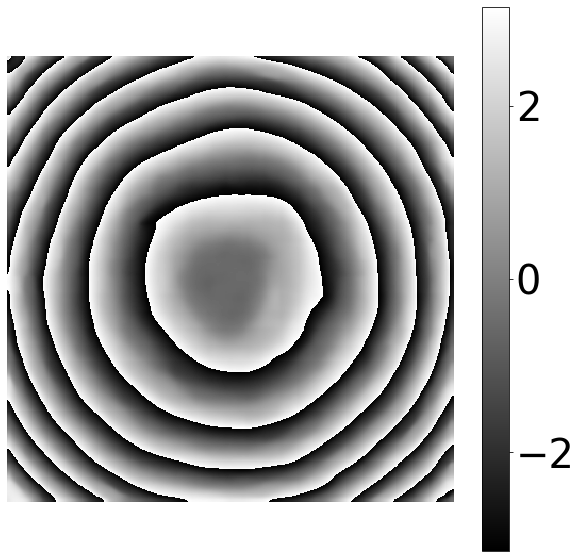}
			\caption{$\delta =1.041 rad$}
			\label{fig:tse_dnn}	
		\end{subfigure}&
		\begin{subfigure}[ht]{0.15\linewidth}
			\includegraphics[width=1\linewidth]{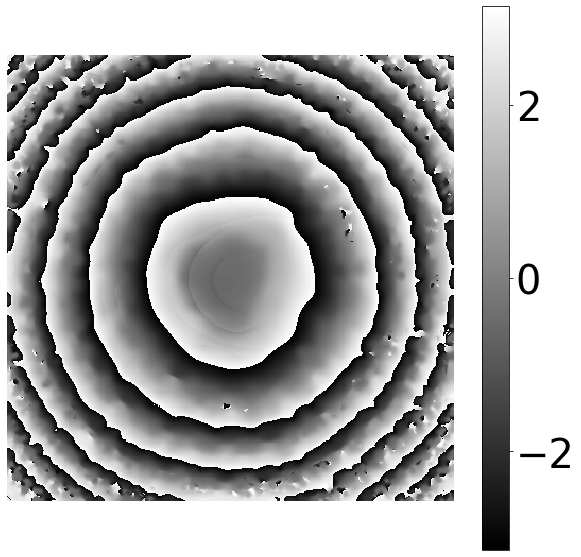}
			\caption{$\delta = 1.051 rad$}
			\label{fig:tse_hht}	
		\end{subfigure}\\
		
		SLEF--RE&
		\begin{subfigure}[ht]{0.15\linewidth}
			\includegraphics[width=1\linewidth]{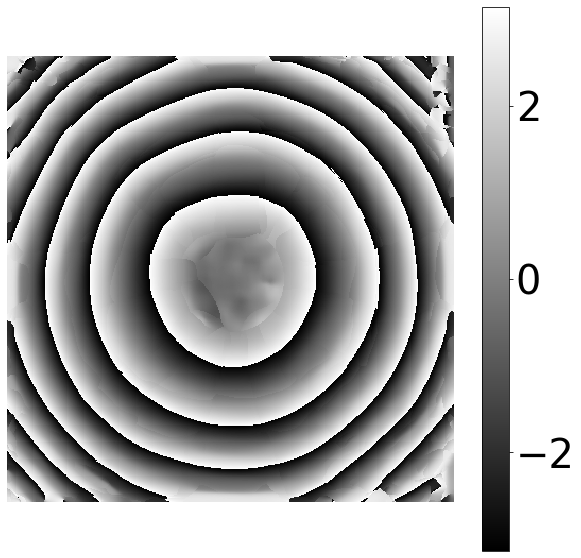}
			\caption{$\delta = 1.171 rad$}
			\label{fig:slef_gfb}	
		\end{subfigure}&
		\begin{subfigure}[ht]{0.15\linewidth}
			\includegraphics[width=1\linewidth]{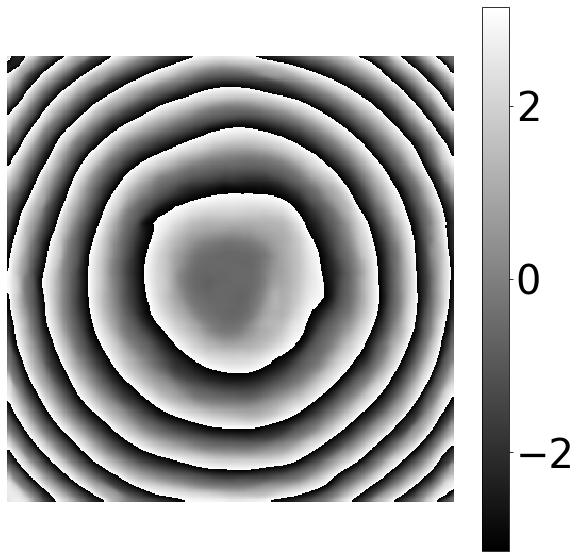}
			\caption{$\delta =1.127 rad$}
			\label{fig:slef_dnn}	
		\end{subfigure}&
		\begin{subfigure}[ht]{0.15\linewidth}
			\includegraphics[width=1\linewidth]{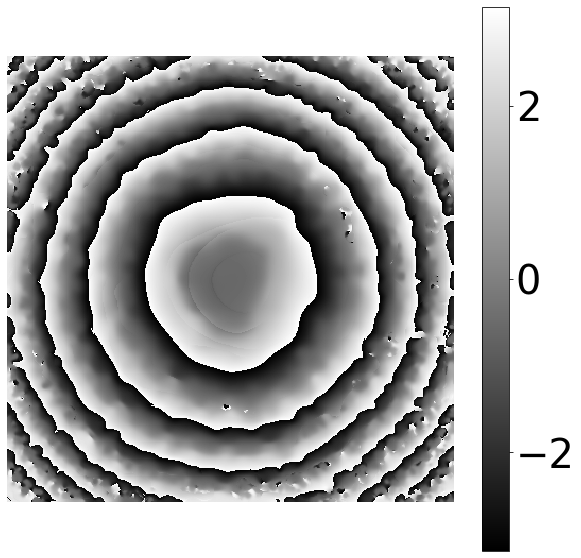}
			\caption{$\delta = 1.164 rad$}
			\label{fig:slef_hht}	
		\end{subfigure}\\
		
		Kreis&
		\begin{subfigure}[ht]{0.15\linewidth}
			\includegraphics[width=1\linewidth]{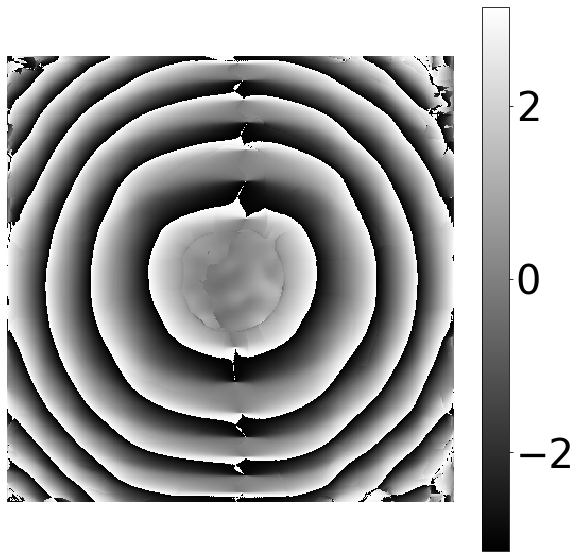}
			\caption{$\delta =1.084 rad$}
			\label{fig:kreis_gfb}	
		\end{subfigure}&
		\begin{subfigure}[ht]{0.15\linewidth}
			\includegraphics[width=1\linewidth]{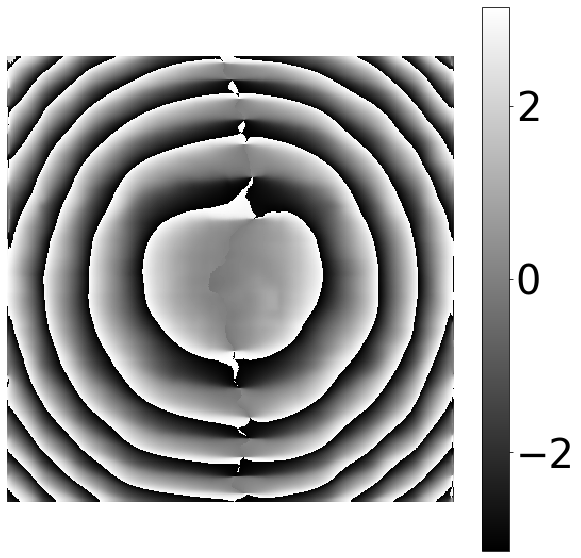}
			\caption{$\delta =1.039 rad$}
			\label{fig:kreis_dnn}	
		\end{subfigure}&
		\begin{subfigure}[ht]{0.15\linewidth}
			\includegraphics[width=1\linewidth]{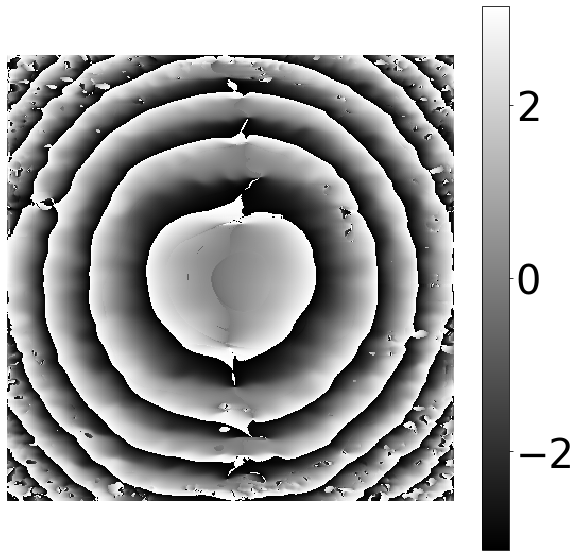}
			\caption{$\delta = 1.031 rad$}
			\label{fig:kreis_hht}	
		\end{subfigure}\\
		
		DDV&
		\begin{subfigure}[ht]{0.15\linewidth}
			\includegraphics[width=1\linewidth]{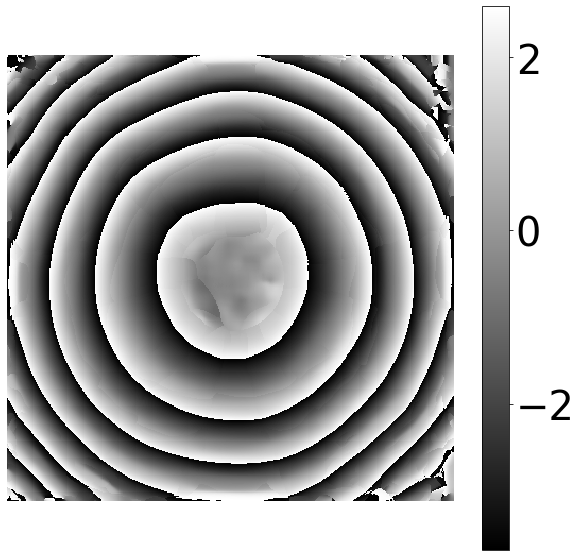}
			\caption{$\delta =1.106 rad$}
			\label{fig:ddv_gfb}	
		\end{subfigure}&
		\begin{subfigure}[ht]{0.15\linewidth}
			\includegraphics[width=1\linewidth]{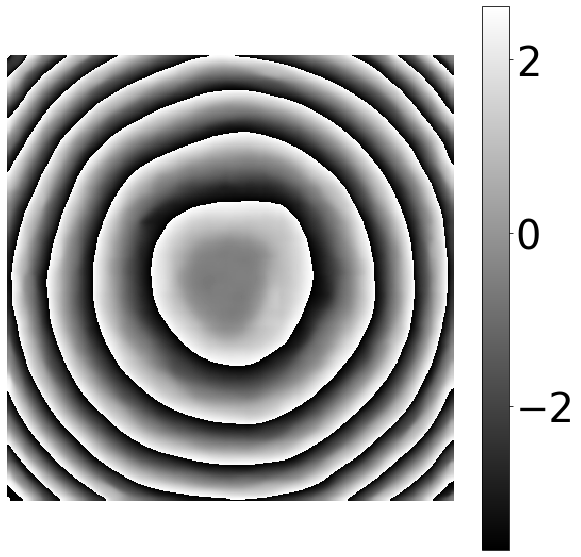}
			\caption{$\delta =1.054 rad$}
			\label{fig:ddv_dnn}	
		\end{subfigure}&
		\begin{subfigure}[ht]{0.15\linewidth}
			\includegraphics[width=1\linewidth]{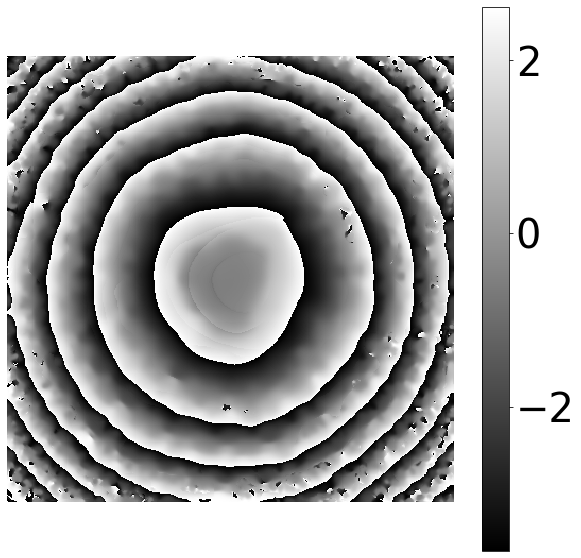}
			\caption{$\delta = 1.058 rad$}
			\label{fig:ddv_hht}	
		\end{subfigure}\\
	\end{tabular}
	\caption{Experimental results part 1.}
	\label{fig:exp_pt1}
\end{figure*}

\begin{figure*}[ht]
	\centering
	\begin{tabular}{lccc}
		\hline
		Algorithm&
		GFB&
		DNN&
		HHT\\
		\hline
		PSC&
		\begin{subfigure}[ht]{0.15\linewidth}
			\includegraphics[width=1\linewidth]{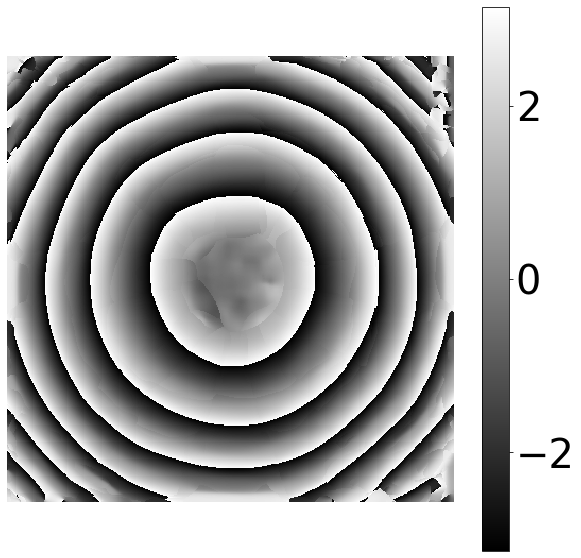}
			\caption{$\delta =1.106rad$}
			\label{fig:psc_gfb}	
		\end{subfigure}&
		\begin{subfigure}[ht]{0.15\linewidth}
			\includegraphics[width=1\linewidth]{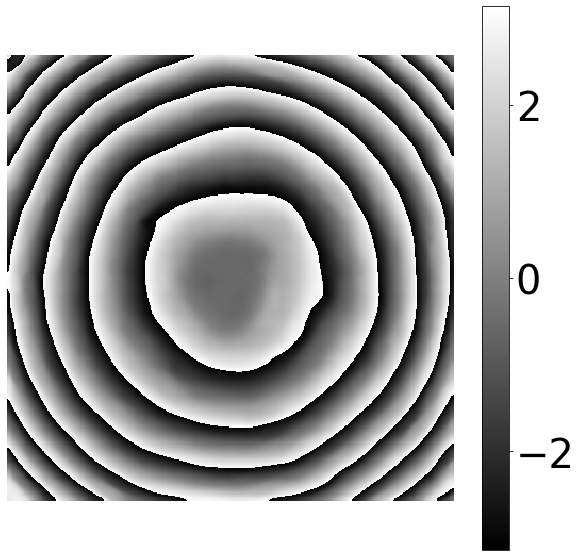}
			\caption{$\delta =1.051 rad$}
			\label{fig:psc_dnn}	
		\end{subfigure}&
		\begin{subfigure}[ht]{0.15\linewidth}
			\includegraphics[width=1\linewidth]{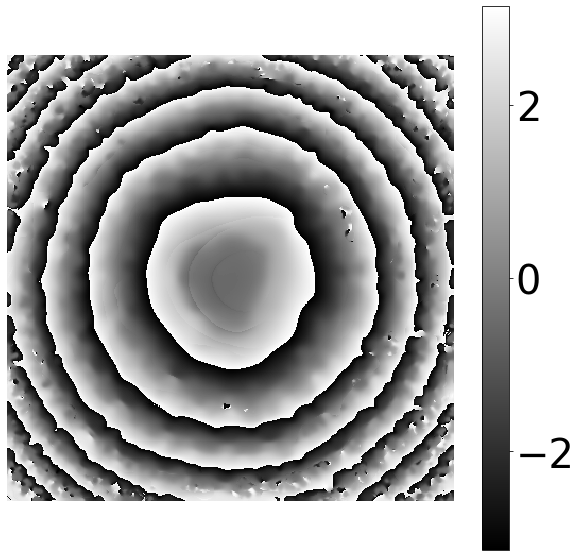}
			\caption{$\delta = 1.06 rad$}
			\label{fig:psc_hht}	
		\end{subfigure}\\
		
		GPSI&
		\begin{subfigure}[ht]{0.15\linewidth}
			\includegraphics[width=1\linewidth]{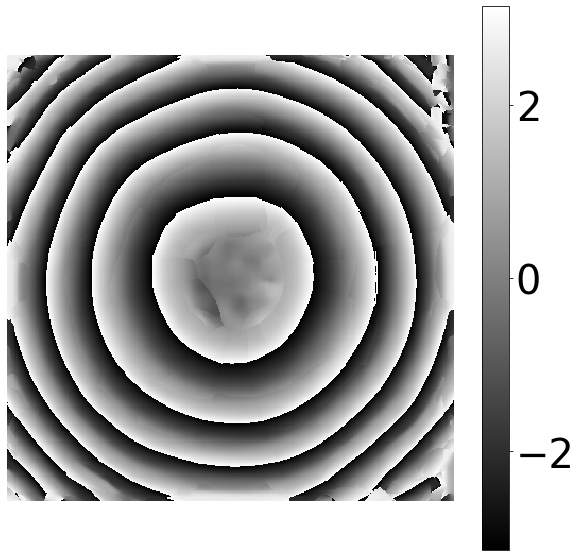}
			\caption{$\delta =0.964rad$}
			\label{fig:gpsi_gfb}	
		\end{subfigure}&
		\begin{subfigure}[ht]{0.15\linewidth}
			\includegraphics[width=1\linewidth]{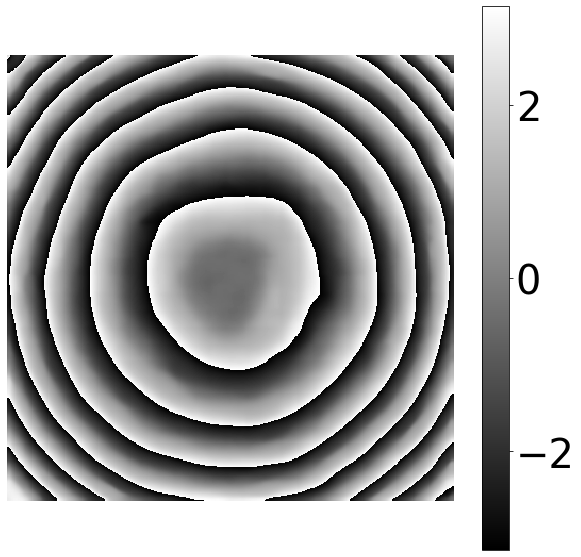}
			\caption{$\delta =0.862 rad$}
			\label{fig:gpsi_dnn}	
		\end{subfigure}&
		\begin{subfigure}[ht]{0.15\linewidth}
			\includegraphics[width=1\linewidth]{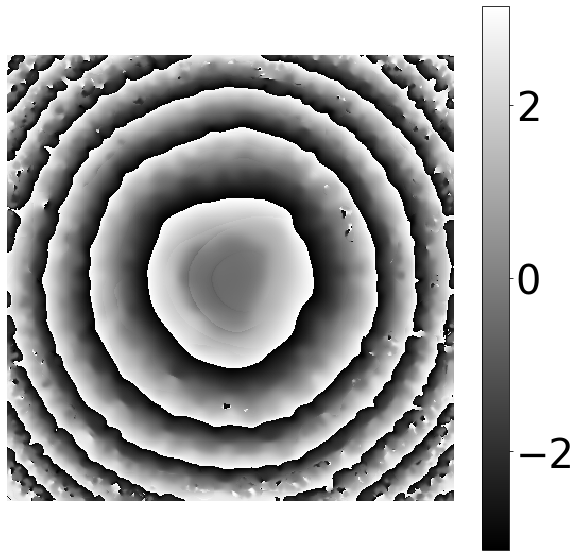}
			\caption{$\delta = 0.975 rad$}
			\label{fig:gpsi_hht}	
		\end{subfigure}\\
		
		GS&
		\begin{subfigure}[ht]{0.15\linewidth}
			\includegraphics[width=1\linewidth]{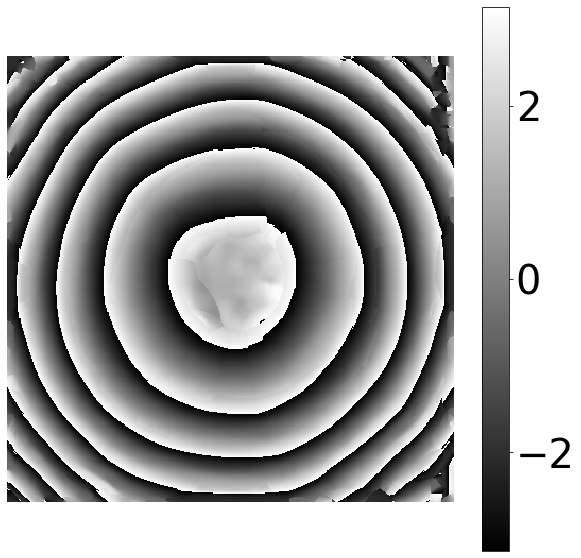}
			\caption{$\delta =1.108rad$}
			\label{fig:gs_gfb}	
		\end{subfigure}&
		\begin{subfigure}[ht]{0.15\linewidth}
			\includegraphics[width=1\linewidth]{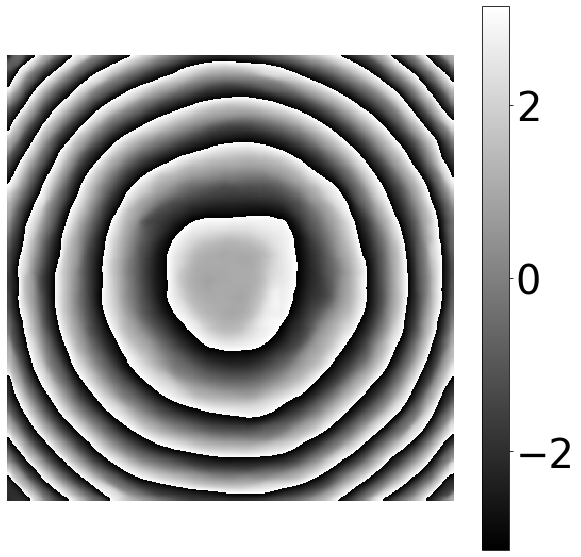}
			\caption{$\delta =1.086 rad$}
			\label{fig:gs_dnn}	
		\end{subfigure}&
		\begin{subfigure}[ht]{0.15\linewidth}
			\includegraphics[width=1\linewidth]{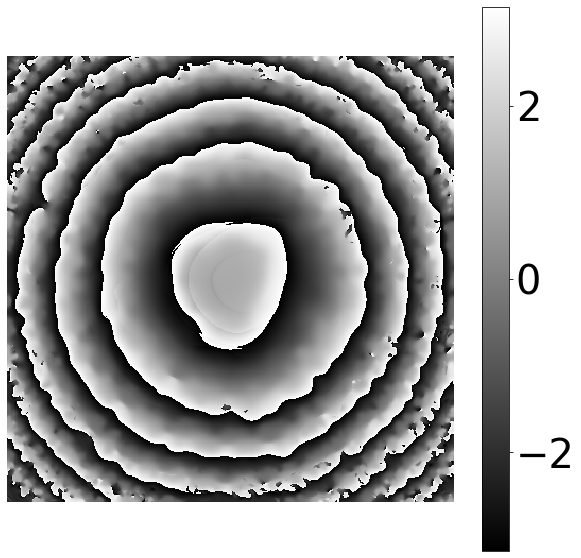}
			\caption{$\delta = 1.060rad$}
			\label{fig:gs_hht}	
		\end{subfigure}\\
		
		EVI&
		\begin{subfigure}[ht]{0.15\linewidth}
			\includegraphics[width=1\linewidth]{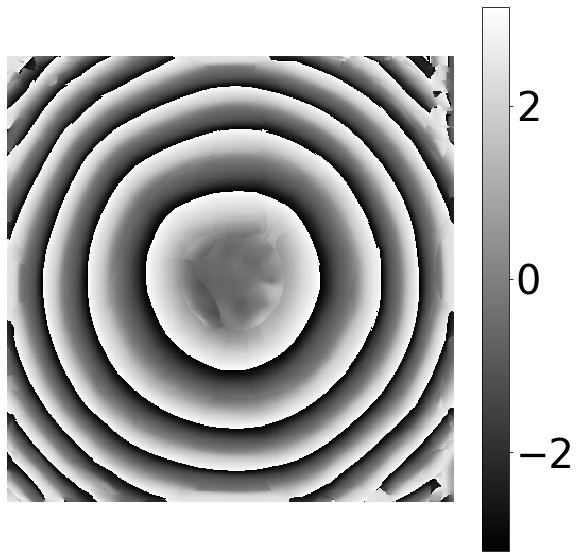}
			\caption{$\delta =1.475rad$}
			\label{fig:evi_gfb}	
		\end{subfigure}&
		\begin{subfigure}[ht]{0.15\linewidth}
			\includegraphics[width=1\linewidth]{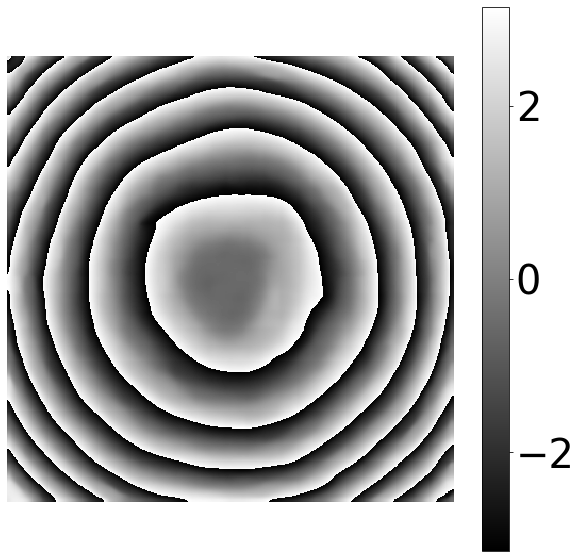}
			\caption{$\delta =1.044 rad$}
			\label{fig:evi_dnn}	
		\end{subfigure}&
		\begin{subfigure}[ht]{0.15\linewidth}
			\includegraphics[width=1\linewidth]{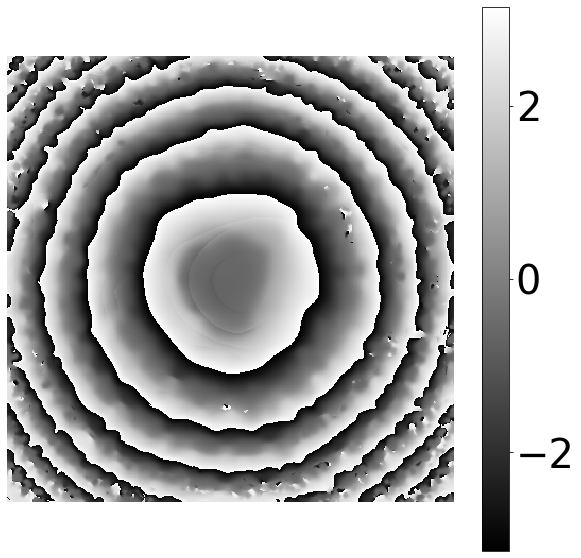}
			\caption{$\delta = 1.289rad$}
			\label{fig:evi_hht}	
		\end{subfigure}\\
		
		RP&
		\begin{subfigure}[ht]{0.15\linewidth}
			\includegraphics[width=1\linewidth]{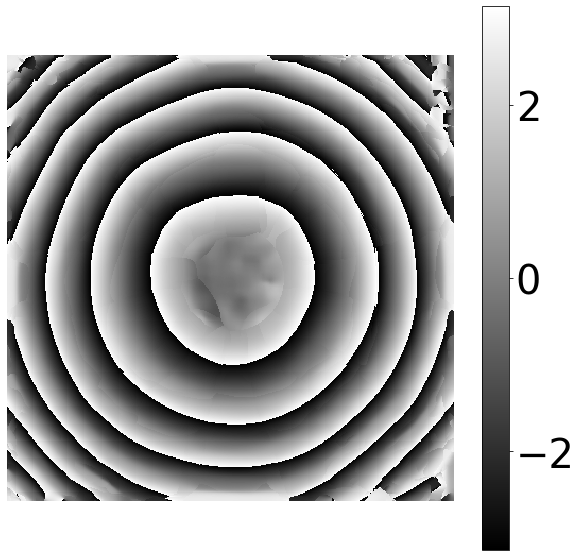}
			\caption{$\delta =1.073rad$}
			\label{fig:rp_gfb}	
		\end{subfigure}&
		\begin{subfigure}[ht]{0.15\linewidth}
			\includegraphics[width=1\linewidth]{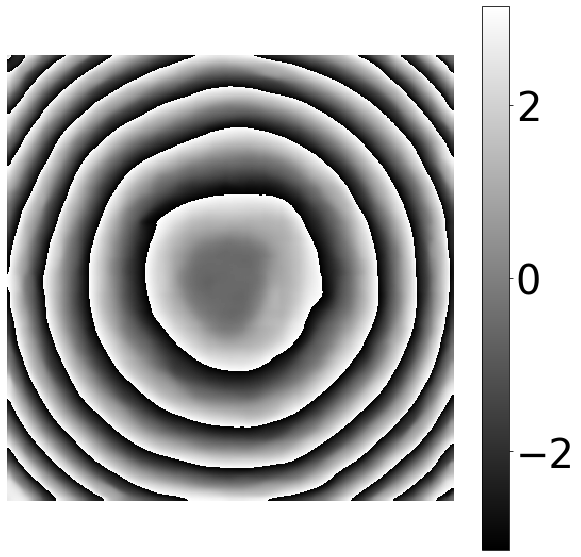}
			\caption{$\delta =1.022rad$}
			\label{fig:rp_dnn}	
		\end{subfigure}&
		\begin{subfigure}[ht]{0.15\linewidth}
			\includegraphics[width=1\linewidth]{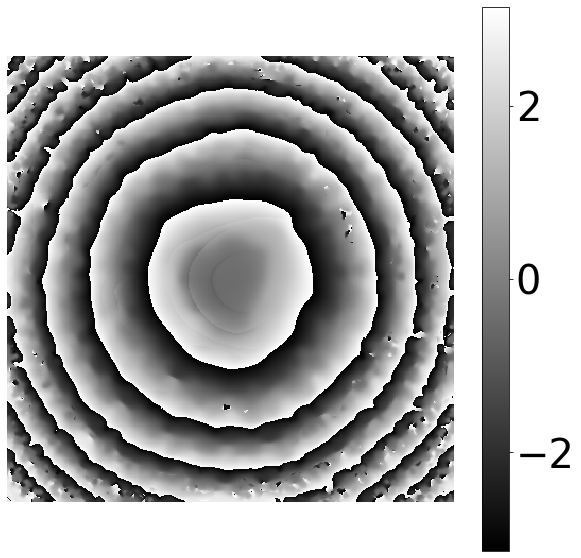}
			\caption{$\delta = 0.947rad$}
			\label{fig:rp_hht}	
		\end{subfigure}\\
		
		QPP&
		\begin{subfigure}[ht]{0.15\linewidth}
			\includegraphics[width=1\linewidth]{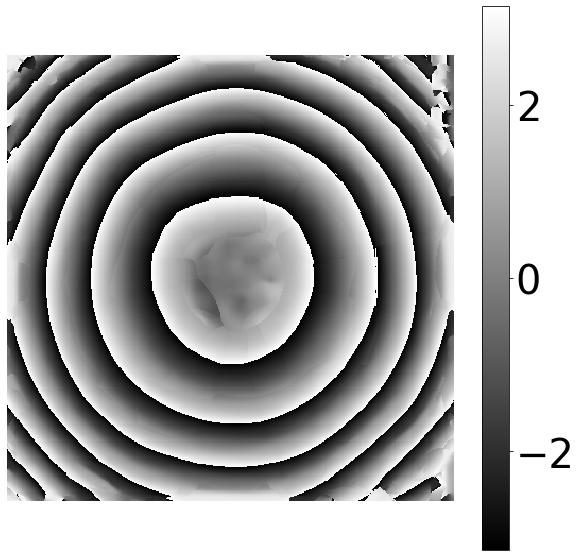}
			\caption{$\delta =1.001rad$}
			\label{fig:qpp_gfb}	
		\end{subfigure}&
		\begin{subfigure}[ht]{0.15\linewidth}
			\includegraphics[width=1\linewidth]{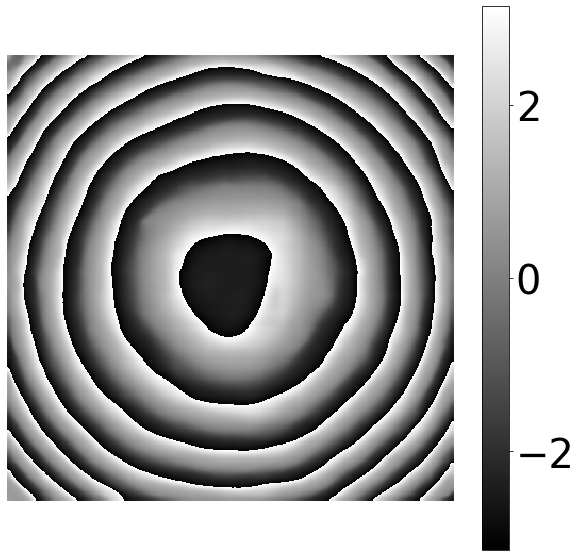}
			\caption{$\delta =4.882rad$}
			\label{fig:qpp_dnn}	
		\end{subfigure}&
		\begin{subfigure}[ht]{0.15\linewidth}
			\includegraphics[width=1\linewidth]{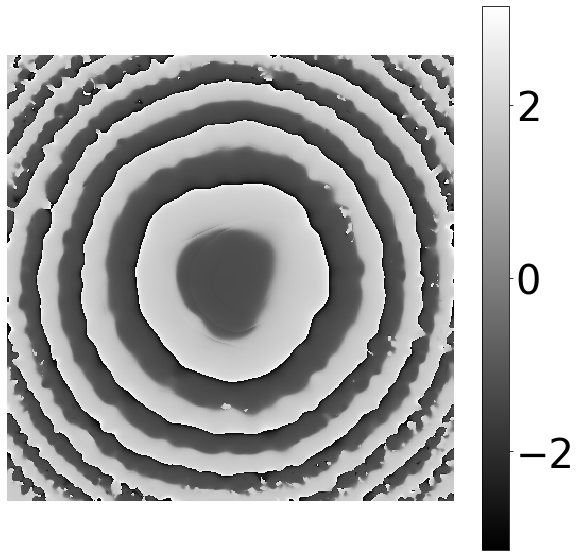}
			\caption{$\delta = 2.555rad$}
			\label{fig:qpp_hht}	
		\end{subfigure}\\

	\end{tabular}
	\caption{Experimental results part 2.}
	\label{fig:exp_pt2}
\end{figure*}

Figures \ref{fig:exp_pt1} and \ref{fig:exp_pt2} present the results of the wrapped phases obtained with the twelve evaluated TS--PSAs by applying the previously mentioned normalization methods.

On each of the phases, we show the estimated phase step calculated by the different TS--PSAs. The most accurate are presented in Figures \ref{fig:mre_gfb}, \ref{fig:ire_gfb}, \ref{fig:tse_hht}, \ref{fig:psc_dnn} and \ref{fig:evi_dnn} with an error less than $0.005 rad$.

On the other hand it can be noticed that those TS--PSAs that presented a wrong estimation, presented high level of detuning, \emph{i.e.} Figures \ref{fig:evi_gfb}, \ref{fig:evi_hht}, \ref{fig:rp_hht}, \ref{fig:qpp_gfb}, \ref{fig:qpp_dnn} and \ref{fig:qpp_hht}. 

The GS algorithm shown in Figures \ref{fig:gs_gfb}, \ref{fig:gs_dnn} and \ref{fig:gs_hht} present a piston term in the phase, due to its orthonormalization process. The presented phase step calculation in such results was obtained by using the algorithm proposed in \cite{flores2019computation}.

The results obtained in Kreis' algorithm (Figures \ref{fig:kreis_gfb}, \ref{fig:kreis_dnn} and \ref{fig:kreis_hht}) as well as DDV (Figures \ref{fig:ddv_gfb}, \ref{fig:ddv_dnn} and \ref{fig:ddv_hht}) were obtained with their original formulation of the phase map. Such algorithms do not use \ref{eq:phi} for the estimation of the phase.

Since the IRE and MRE (Figures \ref{fig:mre_gfb} and \ref{fig:ire_gfb}) algorithms are based on the phases obtained from the GFB, it is not possible to implement them with them directly with the HHT and DNN normalizations, nevertheless, it can be done indirectly by first estimating the phase step with the GFB and then, estimate the phase map by using \eqref{eq:phi} and using the other normalized FPs.

The result presented on Figure \ref{fig:tse_hht} corresponds to the proposal presented in \cite{wielgus2015two}. It can be seen that the algorithm is accurate in the estimation of the phase step, nevertheless, the quality of the phase is noisy due to the HHT. If the pre--filtering process is changed to the GFB or DNNs, the quality of the phase map improves drastically.

As mentioned before, several proposals using HHT as normalization process have been done. Results such as the ones presented in Figures \ref{fig:kreis_hht}, \ref{fig:gs_hht}, \ref{fig:evi_hht} were presented in \cite{saide2017evaluation, trusiak2015two, zhang2019two} respectively.

The RP and SLEF--RE methods were originally proposed with the use of GFB as normalization process, as shown in Figures \ref{fig:rp_gfb} and \ref{fig:slef_gfb} \cite{dalmau2016phase, flores2019robust}. Here, we present their variations by using DNNs and HHT. 

Methods such as DDV, PSC, GPSI, QPP are included given that they present an improvement by using these normalization process and they represent different perspectives of the TS--PSAs.

Finally, QPP presents an inaccurate behavior in the estimation of the phase given that the estimation of the phase step is totally dependent of the chosen window of analysis, which turns out to be non--automatic as the other approaches.

%-------------------------------------------------------------------------------------------------
\section{Discussions}
\label{sec:dis}
%-------------------------------------------------------------------------------------------------
%-----------------------------------------------
\subsection{Kreis and MRE}
%-----------------------------------------------
\noindent Given that the GFBs are capable of delivering the phases ($\psi_1$ and $\psi_2$) and the magnitudes of the interferograms but with the sign ambiguity \cite{rivera2016two} , we use these phases as arguments in \eqref{eq:kreis}, obtaining:

\begin{equation}\label{eq:kreis2}
	\Delta_{RK}(x,y)=
\end{equation}
$$\arctan\Bigg[\frac{\cos \psi_1(x,y) \sin \psi_2(x,y) - \sin \psi_1(x,y) \cos \psi_2(x,y)} {\cos \psi_1(x,y) \cos \psi_2(x,y) + \sin \psi_1(x,y) \sin \psi_2(x,y)}\Bigg],$$
where we use $\Re [\exp z] = \cos(z)$ and $\Im [\exp z] = \sin(z)$.

Then, if we simplify the expression by using the trigonometric properties $\sin(x\pm y)=\sin(x)\cos(y)\pm\cos(x)\sin(y)$ and $\cos(x\pm y)=\cos(x)\cos(y)\mp\sin(x)\sin(y)$, resulting the equation as:

\begin{equation}\label{eq:kreis4}
	\Delta_{RK}(x,y)=\arctan\Bigg[\frac{\sin(\psi_2(x,y)-\psi_1(x,y))} {\cos(\psi_2(x,y)-\psi_1(x,y))}\Bigg].
\end{equation}

Now, if the wrap operator is implemented as $$W(z)=\arctan[\sin(z)/\cos(z)]$$ and $z \in [0,\pi)$, then \eqref{eq:kreis4} is equivalent to the formula \eqref{eq:rivera1}. In other words, one of the oldest TS--PSAs can be as good as a novel algorithm as long as the normalization preprocess is well performed. %That explains the  similar performance of the Rienforced Kreis proposal and the MRE method, with the advantage of the simpler implementation of \eqref{eq:rivera1}.

%-----------------------------------------------
\subsection{QPP variability}
%-----------------------------------------------
\noindent The quadratic phase parameter estimation algorithm for phase demodulation was recently proposed by Kulkarni and Rastogi \cite{kulkarni2018two}. This method is formulated to obtain the parameters of a state space formulation by using the extended Kalman filter. The main disadvantage of this method is that it is completely dependent of the user entries, such as the window to be used to estimate the parameters, the initialization of the state vector and covariance of state estimation error. For this reason we did not include it in the second test of the variable phase shift in Section \ref{sec:shift}, because of the high variability and error that it presented. 

Nevertheless, this method can be automatized by using an Extreme Value detector, as the one we implemented for this study, in order to calculate the initial point of the estimation, so the window to be analyzed can be initialized in a maximum value in order to reduce the variability of the estimation of the parameters in different sets of fringe patterns. This basically consist on generating an EVI map of one of the FPs, and then, generate a random point $p$ where $\forall p \in P$ and $P$ is a set of pixel coordinates with maximum local value.

Normalization methods such as GFBs may not the most suitable for this algorithm because of the probable discontinuities that could be presented or even the HHT normalization could present inconsistencies if the modes are not adequately chosen. However, a DNN normalization with an Extreme Value detection improve drastically their performance (as it was presented in section \ref{ssec:dnnpre}).
%-------------------------------------------------------------------------------------------------
\section{Conclusions}
\label{sec:con}
%-------------------------------------------------------------------------------------------------

\noindent We presented a fair methodology for comparing 12 self-tuning two step phase shift algorithms (TS--PSAs), an active research area nowadays. Most of the analyzed algorithms were recently published and are representative approaches to solve the problem of unknown phase step steps.The performance of the phase step was done in the same framework by using ten different sets of patterns with five different noise levels and five phase shifts. The algorithms were evaluated with and compared using 3 different methods of normalization: GFB, DNN and HHT. Such processes are also representative among the state of the art normalization tools.

The MRE, IRE and TSE algorithms proved to be the more accurate and stable, even though, almost all algorithms presented and error less than $0.05 rad$ at different noise levels as well as the estimation of different phase steps. On the other hand, all the algorithms improve their accuracy as the phase step was closer to $\pi/2$.

The DNN and the HHT normalization processes were applied to most of the algorithms (except for MRE and IRE).The DNN and GFB normalized evaluation presented better accuracy in the phase step estimation in most of the cases with respect to the HHT normalization.  Algorithms such as QPP improved their performance due to the absence of discontinuities and noise that could be generated from the normalization process. In this case, the TSE algorithm had the best performance in calculation of the phase step with the use of the three normalizations. We suggest that the combination of the TSE algorithm is easy and fast to implement given that it present accurate results.

In the experimental part, the results present that the GFB and the DNN normalizations are more robust to speckle noise than the HHT, but in general terms, the normalization method does not have a big influence in the estimation of the phase step with experimental results.

The MRE algorithm was presented in this paper. It had one of the best performances and on the step estimation. We also demonstrated that step calculation formula \eqref{eq:rivera1} is equivalent to \eqref{eq:kreis2}. 

Finally, to answer the question about TS--PSAs: \emph{"Where are we?"}. We concluded that the current formulas for the phase-step estimation produce appropriated results if the FPs that are effectively normalized and de--noised. Hence, the aim in the development of new algorithms should be the improvement of the speed of the normalization algorithms independently of the estimation algorithm.
%-------------------------------------------------------------------------------------------------

\section*{Acknowledgements}
\noindent Author V\'ictor H. Flores thanks Consejo Nacional de Ciencia y Tecnolog\'ia (Conacyt) for the postdoctoral grant provided. This research was supported in part by Conacyt, Mexico (Grant A1-S-43858) and the NVIDIA Academic program.

%-------------------------------------------------------------------------------------------------

\section*{References}

\bibliography{review}

\end{document}